# Data Coding Means and Event Coding Means Multiplexed Over the 100BASE-T1 PMA Sublayer

Alexander Ivanov

*Abstract*—In this paper, we describe a way for enriching the 100BASE-T1 physical layer with an event transfer ability, based on the linguistic multiplexing approach.

## Introduction

FAST ETHERNET, type 100BASE-T1[1] [1] is a relatively new IEEE 802.3 physical layer much inherited from its predecessor, 1000BASE-T, but intended to work over a single unshielded twisted pair. Similar to the predecessor, it consists of just two vertically stacked sublayers, the 100BASE-T1 PCS on its top and the 100BASE-T1 PMA on its bottom.

Interfacing that underlying media, the 100BASE-T1 PMA[2] manipulates on pairs of PAM-3 symbols, therefore we say we employ some means that operates over a transport dictionary consisting of $3^2 = 3 \times 3 = 9$ transport words, two one-of-three possible letters in length each, see Table I, providing the user with a continuous, memory-less, sole-word-wide transfer channel sending one-of-nine word once a time period.

Similar to the 100BASE-T1 PCS[3] that re-arranges the bits between the 4-bit nibbles on the service interface it provides on its top, i.e., the Media Independent Interface, and the 3-bit tuples on the intermediate interface it interacts with the PMA across, we operate on $9 \times 4 = 2 \times 6 \times 3 = 36$-bit quantities collected over a sequence of nine consecutive nibbles to make up a pair of consecutive runs of six consecutive tuples each, see Table II, so solving the rate reconciliation problem.

In the rest of this paper, we use the linguistic multiplexing approach as it was developed up to in [2] as well as the base-prime scrambling concept as it is described in [3], leaving just one, see Table III, thing out of our current scope, assuming it somehow achievable in a reliable degree.

A manuscript of this work was submitted to IEEE Communications Letters November 26, 2022 and rejected as not being in the scope of the journal.

Please sorry for the author has no time to find this work a new home, peer reviewed or not, except of arXiv, and just hopes there it meets its reader, one or maybe various, whom the author beforehand thanks for their regard.

A. Ivanov is with JSC Continuum, Yaroslavl, the Russian Federation.

Digital Object Identifier 10.48550/arXiv.yymm.nnnnn (bundle).

---

[1]Physical Layer for 100 Mbps networking, initially introduced as BroadR-Reach and then standardized in [1], operates in harsh environments, such as of automotive (in-vehicle) or industrial (in-cubicle) applications.

[2]Physical Medium Attachment sublayer, specified in [1], too, implements the 100BASE-T1 Medium Dependent Interface (MDI) transmitting three-level PAM symbols at a rate of $2/3 \times 100 = 66\frac{2}{3}$ million symbols/s.

[3]Physical Coding Sublayer, again specified in [1], implements the service interface (MII) operating at a rate of $1/4 \times 100 = 25$ million nibbles/s, and therefore reconciliates between the PMA (MDI) and its own (MII) rates.

TABLE I
WORDS DEFINITIONS

| Set | Ref. | Card. | Words Within | Signaling |
|---|---|---|---|---|
| echo | □ | 9 | $e_0, e_1, e_2, e_3, e_4, e_5, e_6, e_7, e_8$ | pair of PAM-3 symbols |

NOTE – Implementing scrambling, certain serial images can be assigned arbitrarily.

TABLE II
BIT GROUPS IN A SUPER GROUP

| Size | Even MII#1+4 / PMA@1+6 Groups | | | | Spare | Odd MII#6+9 / PMA@7+12 Groups | | | |
|---|---|---|---|---|---|---|---|---|---|
| 4-bit | #1 | #2 | #3 | #4 | #5 | #6 | #7 | #8 | #9 |
| ↕ regrouping | ○○○○ | ○○○○ | ○○○○ | ○○○○ | ○○○○ | ○○○○ | ○○○○ | ○○○○ | ○○○○ |
| 3-bit | @1 | @2 | @3 | @4 | @5 | @6 | @7 | @8 | @9 @10 @11 @12 |
| time→ | Even Half / Inter-Group Disposition | | | | | Odd Half / Inter-Group Disposition | | | |

TABLE III
OUT-OF-BAND (OoB) SIGNALING

| Set | Ref. | Signal Within | Duration | Signaling |
|---|---|---|---|---|
| — | — | ◊◊◊◊◊ | 5 words = 15 bits | under further studying |

NOTE – To be robust, OoB may use a simpler modulation, e.g., PAM-2 or (D)ME.

TABLE IV
ECHO MULTIPLEXING PROCESS

| Comparison Parameter | input echo | Native Echo Round | output echo | input echo | Forced Echo Round | output echo |
|---|---|---|---|---|---|---|
| Structure | $\frac{1}{2^k}$ → | half's data modulus $2^1 \cdot 8\cdot 8\cdot 8\cdot 8\cdot 8 /2^0$ | → $\frac{1}{2^{k-1}}$ | $\frac{1}{2^k}$ → | half's data modulus $12\cdot 8\cdot 8\cdot 8\cdot 8\cdot 8 /2^9$ → | $\frac{1}{2^{k+9}}$ |
| Technique | | REPRESENTATION | ("< input") | | REPRESENTATION | ("> input") |
| Duration (□ = $e_i$) | | ☐☐☐☐☐☐ six word periods | = 6×3 information bit times = | | ☐☐☐☐☐☐ six word periods | = 180 ns |
| Echo Pool | | $2 \cdot 8^6 = 524{,}288$ samples required | | | $12 \cdot 8^3 = 6{,}144$ samples required | |

## Bit Groups

We distinguish the four even + one spare + four odd MII bit groups (4-bit-long each), the six even + six odd PMA bit groups (3-bit-long each), and the one even + one odd halves (18-bit-long each) in a sequence of 36 consecutive information bits, all inter-arranged the respective way within the latter one we refer to as a super group, see Table II.

We assume that a delimiter, start-of-stream (SSD) or end-of-stream (ESD), corresponds with a robust signaling,[4] maybe out-of-band, see Table III, anyway taking place in the period of the earliest five even PMA bit groups of a super group.

---

[4]100BASE-T1 uses a comparatively long run of zero PAM-3 symbols.

TABLE V  
STREAM CODING SCHEME

| prev super group | #1 | #2 | #3 | #4 | #5 | #6 | #7 | #8 | #9 | next super group | Scenarios, Comments, Notes, Explanations |
|---|---|---|---|---|---|---|---|---|---|---|---|
| idle | data | data | data | data | data | data | data | data | data | data | **Start-of-Stream Delimiter (SSD)** |

Uses OoB signaling and one regular word, always placed in the even half of the current super group comprising its bit groups, to express the start of the transmitted frame and points on the first bit group (it is a MII bit group) related to it, within the stream, respectively. It simply drops all the current echo content and further fills it with the flag bit of the corresponding value followed by the other (if any) information bits replaced by that substitution, ensuring the flag bit encoded/decoded first during the next earliest round of the (native or forced) echo multiplexing process, to clarify the reason of that delimitation as soon as possible:

| Delimiter | Reason = flag bit value |
|---|---|
| SSD | *idle* to *data* transition |
| ESD | *data* to *idle* transition |

SSD/ESD shows no difference except (the value of) the flag bit inserted into the stream at the transmitting side and extracted from that stream at the receiving side, in the form of an extra information bit going alongside the even half with SSD/ESD, to prevent stream inversion mistakes.

Row details:
- any echo, flag bit + 4½×4=18 info bits, $1/2^{19}$, $m_0$
- any echo, flag bit + ½×4=2 info bits, $1/2^{3}$, $m_4$
- any echo, flag bit + no info bits, $1/2^{1}$, $m_8$

**Data or Idle when $E = 2^k > 1$**

When the echo is not cancelled before a half of the current super group, i.e., the half indicates a non-unity data modulus ($E>1$) of the echo before that half occurs, it naturally causes a round of the native echo multiplexing process to occur during that half, resulting in a sample of the native echo pool, that represents (codes) one information bit over what the half itself comprises, thus shortening the echo by a bit per a round and a round per a half. After the half, it therefore indicates the modulus lowered twice, and so on, and so on, until the echo has no information bits at all.

$1/2^k$ minus one echo bit $1/2^{k-1}$

**Data or Idle when $E = 2^0 = 1$**

When the echo is fully cancelled before a half of the current super group, i.e., the half indicates a unity data modulus ($E=1$) of the echo before that half occurs, we artificially inserts an auxiliary bit into the echo, causing a round of the native echo multiplexing process to occur immediately and during that half, fully canceling this new single-bit echo. After the half, it again indicates a unity data modulus and, thus, we can repeat such the insertion procedure, forming an auxiliary, variable bit rate channel whose content is transmitted alongside the MII data and/or event position.

$1/1 \rightarrow 1/2$ minus the aux bit $1/1 \rightarrow \ldots$
auxiliary bit insertion

**Event Position Multiplexing**

Uses a sample of the forced echo pool, always placed in the odd half of the current super group, to code the relative position of the event within that group, pointing on the bit group (PMA bit group) the position of that event corresponds to. Alongside the position, the sample also represents nine information bits, thus enlarging the echo by only nine bits all applied after the newest bits of its current content, if any. Because OoB signaling always occurs in the even half of a super group, it never intersects with any possible round of the forced echo multiplexing process.

$1/2^k$ event position @1÷12 $1/2^{k+9}$

1 or 2 or 3 or 4 or 5 or 6 or 7 or 8 or 9 or 10 or 11 or 12

relative position of the event within the current super group (one selectable among 12 possible positions)

**End-of-Stream Delimiter (ESD)**

| data | data | data | data | data | data | data | data | data | idle | idle | idle | idle | idle |

Uses OoB signaling and one regular word, all placed in the even half of the super group following the current one, to express the end of the transmitted frame and points on the last bit group (MII bit group) related to it, within the stream, respectively. It also inserts the flag bit of the corresponding value into the echo content, before its oldest bit(s), if any, making that bit encoded first during the next earliest round of the echo multiplexing process.

$1/E$ flag bit + no info bits $1/2E$, $m_0$  
$1/E$ flag bit + no info bits $1/2E$, $m_8$

time→ @1 @2 @3 @4 @5 @6 @7 @8 @9 @10 @11 @12     PMA bit groups of the next super group . . .

TABLE VI
STREAM SCRAMBLING SCHEME

| ↓Parameter, Object → | Inversion | Root | Affix | Summary | Details, Comments, Notes, Explanations |
|---|---|---|---|---|---|
| Operation Necessity | optional | mandatory | mandatory | | **Structure** |
| Operation Time and Rate | running continuously, once per a half / echo round | | | $1/f_S$ = 180 ns | The scrambler is composite. It consists of twelve (either eleven) base-2 sub-scramblers and exact one base-259 sub-scrambler, corresponding to the factors of the required echo pool size: |
| Sub-Scrambler Site | 1 × base-2 | 1 × base-259 | 11 × base-2 | | |
| Random Values Needed | no or 1 bit | ≥9 bits | 11 bits | Σ ≥ 21(20) bits | |

NOTE – We mainly focus on the base-259 sub-scrambler while the rest are well-known base-2 ones.

$N_Q$ = 524,288+6,144 = 530,432 = $37^1 \cdot 7^1 \cdot 2^{11}$ = 259$\cdot 2^{11}$

| Protocol | 1000BASE-T | 100BASE-T1 | Proposed |
|---|---|---|---|
| PRNG Output Values | $Sy_n$[3:0] $Sx_n$[3:0] $Sg_n$[3:0] | $Sy_n$[2:0] (+ $Sx_n$[1]) in Idle only | available are up to 6 times by up to 12 values |
| Usage per period | 4+4+4=12 bits per a word | 3(4) out of 12 bits per a word | up to 6·12=72 bits per an echo round |

**Random Number Generator**

100BASE-T1 leverages the 1000BASE-T pseudo random number generator (PRNG) built around a 33-bit-long feedback shift register advancing once per a word and therefore featuring a repetition period of $2^{33}-1$ words, that equals to about 257 s. This predefines the practicability limit when PRNG is updated once per a word, but if it is done once per twice that period, the limit could be doubled to about 515 s, and so on, depending on the number of the PRNG output values demanded per an echo round:

| Demand at Round, $t$, bits | 72 | … | 36 | … | 29 | 28 | 27 | 26 | 25 | 24 | 23 | 22 | 21 | 20 |
|---|---|---|---|---|---|---|---|---|---|---|---|---|---|---|
| Avaliable for Root, $r$, bits | 60(61) | … | 24(25) | … | 17(18) | 16(17) | 15(16) | 14(15) | 13(14) | 12(13) | 11(12) | 10(11) | 9(10) | (9) |
| Repetition Period, $T(t)$, s | ~257 | … | ~515 | … | ~639 | ~662 | ~687 | ~713 | ~742 | ~773 | ~806 | ~843 | ~883 | ~927 |

NOTE – General rule for $t$ or $r$ is:
$T(1 \le t \le 72) = {}^{t}/_{72} \times T(72)$
$T(r) = T(t')$ where $t' = r + 12(11)$

**Plain to Cipher**

$B_{6n} \rightarrow [+] \rightarrow C_{6n}$, with $S_{6n}$ input

**Cipher to Plain**

$C_{6n} \rightarrow [-] \rightarrow B_{6n}$, with $S_{6n}$ input

NOTE – $B_{6n}$, $C_{6n}$, $S_{6n}$ are modulo-259 items.

**Base-259 Sub-Scrambler, Forward and Backward**

It performs a linear (e.g., modulo-259 arithmetic, tabular lookup) operation called "addition" ("+") or its opposite equivalent called "subtraction" ("−"), once per each echo round, i.e., every six words, to scramble and, respectively, de-scramble the root part of the code point (that is a modulo-$N_Q$ item) representing the result of the current echo multiplexing round so:

$C = S(B)$, $B = S^{-1}(C)$, $S^{-1}[S(\text{not } B)] \ne B$, $S[S^{-1}(\text{not } C)] \ne C$.

| | Symmetric Solutions Targeting $|\Delta x|$ = 1 | | | | | Symmetry | | Probability Distribution Intervals | | Even/Odd Ratio | | Practicability | |
|---|---|---|---|---|---|---|---|---|---|---|---|---|---|
| $r$, bits | $m_{even}$ | $m_{odd}$ | $|\Delta m|$ | $x_{even} = x_{odd} + \Delta x$ | $x_{odd}$ | $\Delta x$ | "e" | "o" | central | side even-to-odd ratio | "core" | "leaf" | Unbalance | Observation Time |
| 9 | 253 | 6 | 247 | 2 > 1 | | +1 | yes | — | $x_{even}$ | $x_{even} : x_{odd}$ = 42 : 1 | 1 : 0 | 42 : 1 | +1.172/−49.41 % | ~0.1 ms << $T(r)$ |
| 27 | 43 | 216 | 173 | 518,216 > …,215 | | +1 | — | yes | $x_{even}$ | $x_{even} : x_{odd}$ = 21 : 108 | 2 : 43 | 1 : 5 | +1.609/−0.320 ppM | ~24.2 s < $T(r)$ |
| 28 | 173 | 86 | 87 | 1,036,430 < …,431 | | −1 | yes | — | $x_{even}$ | $x_{even} : x_{odd}$ = 2 : 1 | 1 : 0 | 2 : 1 | +0.644/−0.320 ppM | ~48.3 s < $T(r)$ |
| 29 | 87 | 172 | 85 | 2,072,860 < …,861 | | −1 | yes | — | $x_{even}$ | $x_{even} : x_{odd}$ = 1 : 2 | 1 : 0 | 1 : 2 | +0.162/−0.320 ppM | ~96.3 s < $T(r)$ |
| 35 | 129 | 130 | 1 | 132,663,082 < …,083 | | −1 | — | yes | $x_{even}$ | $x_{even} : x_{odd}$ = 1 : 1 | 1 : 2 | 1 : 1 | +3.754/−3.783 ppB | ~1.7 h >> $T(r)$ |
| 36 | 1 | 258 | 257 | 265,326,166 > …,165 | | +1 | — | yes | $x_{even}$ | $x_{even} : x_{odd}$ = 0 : 129 | 1 : 0 | 0 : 129 | +3.754/−0.015 ppB | ~3.4 h >> $T(r)$ |
| 37 | 257 | 2 | 255 | 530,652,330 < …,331 | | −1 | yes | yes | $x_{even}$ | $x_{even} : x_{odd}$ = 128 : 1 | 1 : 0 | 128 : 1 | +1.870/−0.015 ppB | ~6.9 h >> $T(r)$ |
| 38 | 255 | 4 | 251 | 1,061,304,660 < …,661 | | −1 | — | yes | $x_{even}$ | $x_{even} : x_{odd}$ = 127 : 2 | 85 : 2 | 85 : 1 | +0.928/−0.015 ppB | ~13.7 h >> $T(r)$ |
| 44 | 3 | 256 | 253 | 67,923,498,240 < …,241 | | −1 | yes | yes | $x_{even}$ | $x_{even} : x_{odd}$ = 1 : 128 | 1 : 0 | 1 : 128 | +0.000/−0.015 ppB | ~36.7 d >> $T(r)$ |
| 45 | 253 | 6 | 247 | 135,846,996,482 > …,481 | | +1 | yes | — | $x_{even}$ | $x_{even} : x_{odd}$ = 42 : 1 | 1 : 0 | 42 : 1 | +0.000/−0.007 ppB | ~73.3 d >> $T(r)$ |

NOTE – Among considered $\log_2 259 < 9 \le r \le 60(61) = 72-12(11)$, cases for $r$ = 10…26, 30…34, 39…43, 46…60, (61) have no symmetric solution, thus, they are not shown.

We search for a solution of the following system:

$m_{even} \cdot x_{even} + m_{odd} \cdot x_{odd} = 2^r$
$m_{even} + m_{odd} = 259$

where $m_{even}, m_{odd}, x_{even}, x_{odd} \ge 0$
$(x_{even} + x_{odd}) \mod 2 \ne 0$

NOTE – Indeces "even" and "odd" are in respect to the values of $x$'s.

Among the solutions, the symmetric ones are more easy to implement:

Criterion "e" : $\max\{m_{even}-1, m_{odd}\} \mod \min\{m_{even}-1, m_{odd}\} = 0$
Criterion "o" : $\max\{m_{even}, m_{odd}-1\} \mod \min\{m_{even}, m_{odd}-1\} = 0$

Thus, we consider $r$ = 15 the optimal choice while its unbalance negligible.

| | Solutions Targeting $|\Delta m|$ = 1, $|\Delta x| \rightarrow$ min | | | Practicability | |
|---|---|---|---|---|---|
| $r$, bits | $x_{even}$ | $x_{odd}$ | $|\Delta x|$ | Unbalance | Observation Time |
| 14 | 126 | 1 | 125 | +99.18/−98.41 % | ~2.9 ms << $T(r)$ |
| 15 | 122 | 131 | 9 | +3.543/−3.571 % | ~5.9 ms << $T(r)$ |
| 19 | 2,082 | 1,967 | 115 | +2.851/−2.830 % | ~0.1 s < $T(r)$ |
| 20 | 4,034 | 4,063 | 29 | +0.357/−0.360 % | ~0.2 s < $T(r)$ |
| 23 | 32,402 | 32,375 | 27 | +0.042/−0.042 % | ~1.5 s < $T(r)$ |
| 26 | 259,086 | 259,129 | 43 | +0.008/−0.008 % | ~12.1 s < $T(r)$ |

NOTE – Cases for $r$ = 9…13, 16…18, 21, 22, 24, 25, 27 have no solution, ≥28 may have one.

**Randomity Base Conversion**

It picks up a single point in a 259-point [Base-259] Space, based on $r$ values forming a $2^{r \ge 9}$-point [Base-$2^{r \ge 9}$] Space, which are sourced by PRNG.

| Relationship Parameters, $r$ = 15 | left "leaf" (neg.) | | | | "core" | | | right "leaf" (pos.) | | | |
|---|---|---|---|---|---|---|---|---|---|---|---|
| Probability Distribution Intervals (Bins) | {001} | {002} | … | {127} | {128} | {129} | {130} | {131} | {132} | {133} | … | {258} | {259} |
| Number of Points in Bin, *always one* Prob. per Point, *quasi-uniform*, %, ~ | 1 .400 | 1 .373 | … | 1 .400 | 1 .373 | 1 .400 | 1 .373 | 1 .400 | 1 .373 | 1 .400 | … | 1 .373 | 1 .400 |
| Probability per Point, uniform = $1/2^r$ Number of Points in Bin, $x_{even}$ or $x_{odd}$ | $1/2^{15}$ 131 | $1/2^{15}$ 122 | … | $1/2^{15}$ 131 | $1/2^{15}$ 122 | $1/2^{15}$ 131 | $1/2^{15}$ 122 | $1/2^{15}$ 131 | $1/2^{15}$ 122 | $1/2^{15}$ 131 | … | $1/2^{15}$ 122 | $1/2^{15}$ 131 |
| Probability Distribution Intervals (Bins) | −129 | −128 | … | −3 | −2 | −1 | central | +1 | +2 | +3 | … | +128 | +129 |

TABLE VII
STREAM MAPPING SCHEME

| Echo Samples Selection Criteria | | | Resulting Subset | |
|---|---|---|---|---|
| Longest Droop | DC Unb. Bound | Transits Count | Samples Within | App. |
| ≤ 7 or ≤ 7 or | ± 7 or | ≥ 3 | various numbers | < $N_Q$ |
| ≤ 8 / ≤ 8 | ± 8 | ≥ 2 | 530,432 | = $N_Q$ |
| > 8 / > 8 | ± 12 | ≥ 0 | up to $9^6$ = 531,441 | > $N_Q$ |

NOTE – Droop, in PAM-3 symbol time periods, is for the head / tail of an image.
NOTE – DC unbalance bound, in unit dozes, is for any serial image in the subset.
NOTE – Transits count, in PAM-3 level jumps, is per a serial image in the subset.

TABLE VIII
KEY FEATURES OF THE PROPOSED WAY

| Parameter | Description |
|---|---|
| Coding Method, Technique | linguistic multiplexing, echo representation only |
| Coding Delay—Transmit | 12 words = 360 ns = 100BASE-T1 TX delay limit |
| Coding Delay—Receive | 24 words = 720 ns < 100BASE-T1 RX delay limit |
| Event Fixation Resolution | up to one event per a word = 30 ns [uncertainty ±15 ns] |
| Event Transmission Rate | up to one event per 60 words ≈ 0.56 Mev/s |
| Physical (PMA) Line Code | three-level pulse-amplitude modulation (PAM-3) |
| Avg. (Max) Droop Duration | about 3 (16) PAM-3 symbols = about 45 (240) ns |
| Avg. Stream DC Unbalance | theoretically, zero over an infinite integration time |
| Avg. Number of Transitions | about 7.3 jumps per every 12 PAM-3 symbols |

## ECHO REPETITION MEANS

Following the synthesis flow developed in [2], we define the necessary echo multiplexing process the way we assume the most natural, so an echo multiplexing round, native or forced, see Table IV, fits exactly in and begins strictly at a half of a super group, i.e., occupies six consecutive transport words all related with either even or odd, but not of both anyway, PMA bit groups, as a single sample resulted from the round.[5]

As well, we restrict the multiplexing process so a delimiter occurs only during the even half of a super group, in contrast to a forced echo that occurs only during the odd half of a super group, thanks to that, these two never intersect in the stream, and a native echo occurs during any period which is free of a delimiter and a forced echo, repeating the respective process either (when it really is) due to a real necessity, or (otherwise) due to a handy generated cause,[6] see Table V, eliminating the need in any other multiplexing scenarios.

By this, we preserve the behavior of the revised data transfer means, as the user observes it at the service interface, as much similar as possible with the behavior of the original physical layer,[7] and simultaneously introduce an extra means intended to transfer an event, new to the original.

## EVENT FIXATION MEANS

We distinguish up to $2 \times 6 = 12$ positions an event can be associated with, during a super group, in respect to the number of PMA bit groups comprising it, see Table V, that allows for an event fixation means to have a resolution of one event per a transport word and an accuracy of half a word.[8]

As it was described earlier, a super group—during which an event occurs—always conveys a forced echo sample that represents, among other information, the position of the event in that super group, in its odd half.

## BASE-259 SUB-SCRAMBLER

It is the most complicated step on the proposed way and an integral part of the whole scrambler, see Table VI, synthesized following the approach described in [3], with some reasonable modification in pseudo random number generation.

Because the 100BASE-T1 PCS implements an excessive—in the sense of the number of statistically independent binary values generated per a transport word period—PRNG, leveraged from its predecessor, the 1000BASE-T PCS, we use that excessiveness to generate base-259 pseudo random numbers, bringing an additional—however appropriate—inaccuracy but eliminating the need in a biasing anchor used in [3].

Based on this, we define a rule for the necessary base-$2^{>1}$ to base-259 conversion and then search for its optimal variant we finally select to ground the design on. Other details of the base-259 sub-scrambling are similar to the respective features of a general base-prime scrambling introduced in [3].

Compared to a base-2 sub-scrambler who seems trivial, the base-259 one seems very complex, but that complexity enables us to scramble on integral entities, i.e., echo samples, instead of individual bits, keeping the transferred stream consistent as much as possible,[9] see Table VII.

## CONCLUSION

We highlighted a way enriching the 100BASE-T1 physical layer with an event transfer ability, based on the redundancy, which is not so big but enough for, of the underlying transport means provided by its PMA sublayer. We tried to isolate all, or at least most of, the differences inside the PCS.

We hope that this way demonstrates interesting properties, see Table VIII, and, therefore, would be applicable in various mixed communication and synchronization tasks.

---

[5] There are $9^6 = 3^{12} = 531,441$ possible, distinguishable echo samples all together comprising the total echo pool, the native and forced echo pools consisting of $2 \cdot 8^6 = 524,288$ and $12 \cdot 8^3 = 6,144$ samples, respectively, are non-intersecting ($2 \cdot 8^6 + 12 \cdot 8^3 < 9^6$) subsets of which.

[6] In a form of an auxiliary bit value transferred per such a round, not visible to the PHY user at the MII, but still accessible indirectly, if needed.

[7] One difference lies in the implementation of an error propagation function, to convey up to one error per a frame as the 100BASE-T1 PCS does, that is not defined but still achievable in the proposed design, e.g., by truncation or extension of a frame to an odd number of MII nibbles.

[8] It can be used with a practically comfortable limitation, such as only nine positions ($9 < 12_{max}$) in respect to the number of MII bit groups, instead of those of PMA, providing a proportionally coarser resolution of an event per a nibble and a lower accuracy of half the nibble period.

[9] Assuming $N_Q = 259 \cdot 2^{11} = 530,432$, we consider the stream as a series of independent base-$N_Q$ values, each scrambled as a whole, individually, by a modulo-$N_Q$ unambiguously reversible operation, therefore, keeping its base the same both before and after the scrambling, see Tables VI and VII.

# Data Coding Means and Event Coding Means Multiplexed Over a 10BASE-T/Te MAU-like Entity

Alexander Ivanov

*Abstract*—We are seeking a way to supplement the 10BASE-T MAU with an event transfer ability, enabling either the linguistic multiplexing approach or an approach like PCS microframing. The way we suggest and describe in this paper is also applicable for the 10BASE-Te MAU and similar protocols.

*Index Terms*—Ethernet, linguistic multiplexing, multiplexing, like-a-Manchester, LaM, 10BASE-T, 10BASE-Te.

## INTRODUCTION

ORIGINAL-speed Ethernet, type 10BASE-T per IEEE Std 802.3i-1990 [1] and its newer, energy-efficient variant, 10BASE-Te per IEEE Std 802az-2010 [2], both provide for a 10 Mbps Ethernet duplex connection over a dual-pair twisted-pair cable link, based upon a Manchester encoding variant as the line code they both implement to interact. Compared with its higher-speed successors, the corresponding physical layer looks very simple and consists only of the medium attachment unit (MAU) which embodies the respective physical medium attachment (PMA) sublayer as well as its exposed, compatible media dependent interface (MDI).

Inside of the 10BASE-T/Te PMA sublayer, there is no extra block-based or another-buffering coding means like inside of its successors, and therefore, generally speaking, such a PMA sublayer almost directly passes a signal—like CD0 and CD1 presented during transmissions of information bits—between the MDI under the bottom and the physical signaling (PLS) interface at the top of itself, respectively, merely conditioning the passed signal in its shape and power.

In this paper, we describe a way to construct a reciprocally multiplexed but independently operated data coding means and event coding means—in a form of some block-based physical coding sublayer (PCS)—feasible and stackable atop the PMA sublayer of the mentioned MAU, exploiting on the designated signaling transparency of the behavior the MAU demonstrates as well as on the known redundancy of the line code the MAU implements, contemporaneously.

## RETHINKING THE 10BASE-T SIGNALING

Considering the Manchester encoding variant employed in 10BASE-T, see Table I, we conclude that the 10BASE-T MDI signal observed during transmission of information bits, i.e., a

A manuscript of this work was submitted to IEEE Communications Letters November 26, 2022 and rejected as not being in the scope of the journal.

Please sorry for the author has no time to find this work a new home, peer reviewed or not, except of arXiv, and just hopes there it meets its reader, one or maybe various, whom the author beforehand thanks for their regard.

A. Ivanov is with JSC Continuum, Yaroslavl, the Russian Federation.

Digital Object Identifier 10.48550/arXiv.yymm.nnnnn (bundle).

TABLE I
EXAMPLE VALID MANCHESTER ENCODING

| Bit Time Period, $n \rightarrow$ | $n+1$ | $n+2$ | $n+3$ | $n+4$ | $n+5$ | $n+6$ |
|---|---|---|---|---|---|---|
| Data Bit Stream | CD0 | CD0 | CD1 | CD1 | CD1 | CD0 |
| Resulting Waveform | L H | L L | H H | L H | L H | H H |
| Pulse Types Found: | | | | | | |
| • Narrow Pulse ⟨N⟩ | ⟨N⟩ | ⟨N⟩ | | ⟨N⟩ | ⟨N⟩ | ⟨N⟩ |
| • • positive N+ | | N+ | | | N+ | |
| • • negative N− | N− | | | N− | | N− |
| • Wide Pulse ⟨W⟩ | | | ⟨W⟩ | | | ⟨W⟩ |
| • • positive W+ | | | | | | W+ |
| • • negative W− | | | W− | | | |
| Equivalently Represented MDI Pulse Train | N− | N+ | W− | N+ | N+ | N− W+ |
| Duration [BT = one bit time] | BT/2 | BT/2 | BT | BT/2 | BT/2 | BT/2 BT |
| Bit Time Half Period $\rightarrow$ | 2/2 1/2 | 2/2 1/2 | 2/2 1/2 | 2/2 1/2 | 2/2 1/2 | 2/2 1/2 |

TABLE II
PROPOSED LIKE-A-MANCHESTER ENCODING

| Permissible Pulse Types $\rightarrow$ | ⋯ | N+ | N− | W+ | W− | ⋯ |
|---|---|---|---|---|---|---|
| MDI Pulse Width | ⋯ | nar-pos. | -row neg. | wide positive | wide negative | ⋯ |
| MDI Pulse Polarity | | | | | | |
| MDI Pulse Duration [BT = one bit time] | ⋯ | BT/2 | BT/2 | BT | BT | ⋯ |
| Transport Letter Stream Found | ⋯ | J | J | J | K | J | K | J | ⋯ |
| Signal High — Positive Level | | | | H H | | H | | H | |
| Signal Average — Zero Level | | + | + | + + + | + | + − | − | + | |
| Signal Low — Negative Level | | L | L | − − − | | L L | L L | L | |
| Signal Polarities @ Half Periods | | − + | + − | − + + + | + − | − − | − + | |
| DC Bias, accumulated per letter | ⋯ | =0 | =0 | =0 | +1 | =0 | −1 | =0 | ⋯ |
| Letter Time Half Period, within $t$ | ½ ⅔ | ½ ⅔ | ½ ⅔ | ½ ⅔ | ½ ⅔ | ½ ⅔ | ½ ⅔ |
| Letter Time Period, $t$ [t:n=2:1] $\rightarrow$ | ⋯ | $t+1$ | $t+2$ | $t+3$ | $t+4$ | $t+5$ | $t+6$ | $t+7$ | ⋯ |

period when the resulting waveform purely consists of the CD0 and CD1 signal shape repetitions, is sufficiently representable as some pulse train comprising consecutively and continuously running pulses varying in their polarity and duration, i.e., each being positive-or-negative while narrow-or-wide.

In support of our assumption made above, we see that the formal specification given in [1] also defines the MDI signal in the terms of narrow and wide pulses running in negative and positive polarities. Nominally, the duration of a wide pulse is exact one information bit time period, that is one tenth of a microsecond, i.e., 1/10 MHz = 100 ns, while the duration of a narrow pulse is just a half of the duration of a wide one, i.e., 1/20 MHz = 50 ns. Concerning the polarities, it is enough in our scope to say that a positive pulse and a negative pulse are

TABLE III
TINY WORDS

| Possible Serial Images | VALID Serial Images | Balance and Level Inversion | MDI Signal Pattern if the previous period ends up with L | MDI Signal Pattern with H | Transport Purpose when the serial image mask is J... | Transport Purpose ...J |
|---|---|---|---|---|---|---|
| K K K K | | | | | | |
| K K K J | | | | | | |
| K K J K | | | | | | |
| K K J J | | | | | | |
| K J K K | | | | | | |
| K J K J | → K J K J | 2J-2K | L [=0] L | H [=0] H | — | $B_1$ |
| K J J K | | | | | | |
| K J J J | → K J J J | 3J-1K $^{inv}$ | L [−1] H | H [+1] L | — | $B_4$ |
| J K K K | | | | | | |
| J K K J | | | | | | |
| J K J K | → J K J K | 2J-2K | L [=0] L | H [=0] H | $A_1$ | — |
| J K J J | → J K J J | 3J-1K $^{inv}$ | L [+1] H | H [−1] L | $A_2$ | $B_2$ |
| J J K K | | | | | | |
| J J K J | → J J K J | 3J-1K $^{inv}$ | L [−1] H | H [+1] L | $A_3$ | $B_3$ |
| J J J K | → J J J K | 3J-1K $^{inv}$ | L [+1] H | H [−1] L | $A_4$ | — |
| J J J J | → J J J J | 4J-0K | L [=0] L | H [=0] H | $A_5$ | $B_5$ |

NOTE – Tiny words are four-letter = two-bit long, relate with the RMII path width.

TABLE IV
EXAMPLE CODING MEANS BUILT AROUND TINY WORDS

| Input | Multiplexing | Scrambling | Encoding | Serializing | Output |
|---|---|---|---|---|---|
| event load/idle $b_{2n+1}$ $b_{2n+0}$ b's are independent, word size in bits is $m = 2$ | M, mem = multiplexing site memory | single bit delay line = $z^{-1}$, S5 | E, $z^{-1}$ | P/S, $n:t = 1:2$ | $q_t$ |

| Scrambling Out/Result | Encoding Out/Result | Serial Stream | Scrambling Out/Result | Encoding Out/Result | Serial Stream |
|---|---|---|---|---|---|
| $1_{(10)} = 001_{(2)}$ | 1 0 1 0 = $A_1$ | J K J K | $1_{(10)} = 001_{(2)}$ | 0 1 0 1 = $B_1$ | K J K J |
| $2_{(10)} = 010_{(2)}$ | 1 0 1 1 = $A_2$ | J K J J | $2_{(10)} = 010_{(2)}$ | 1 0 1 1 = $B_2$ | J K J J |
| $3_{(10)} = 011_{(2)}$ | 1 1 0 1 = $A_3$ | J J K J | $3_{(10)} = 011_{(2)}$ | 1 1 0 1 = $B_3$ | J J K J |
| $4_{(10)} = 100_{(2)}$ | 1 1 1 0 = $A_4$ | J J J K | $4_{(10)} = 100_{(2)}$ | 0 1 1 1 = $B_4$ | K J J J |
| $5_{(10)} = 101_{(2)}$ | 1 1 1 1 = $A_5$ | J J J J | $5_{(10)} = 101_{(2)}$ | 1 1 1 1 = $B_5$ | J J J J |
| When the previous mask ≠ J...J | | | When the previous mask = J...J | | |

TABLE V
PAGE BALANCING MEANS

| In/Argument | Delay Lines = Memory | Look-up Table | Out/Result |
|---|---|---|---|
| dependent $s > m$ bits, $s > m$, from scrambling | $z^{-1}$ ... $z^{-1}$ $z^{-1}$ (ξ ... 2 1) | E, ξ | dependent $M = 2m$ bits, $M = 2m$, into serializing |

| Memory Depth, ξ | Number of Combinations | Probability of Occurrence Page A | Probability of Occurrence Page B | Unbalance $\Delta p = p_A − p_B$ | % |
|---|---|---|---|---|---|
| ξ = 1 | $3^1 = 3^1 = 3$ | $p_A = 2/3$ | $p_B = 1/3$ | $\Delta p = 1/3$ ≈ | 33.3% |
| ξ = 2 | $3^2 = 3^2 = 9$ | $p_A = 5/9$ | $p_B = 4/9$ | $\Delta p = 1/9$ ≈ | 11.1% |
| ξ = 3 | $3^3 = 3^3 = 27$ | $p_A = 14/27$ | $p_B = 13/27$ | $\Delta p = 1/27$ ≈ | 3.7% |
| ξ = 4 | $3^4 = 3^4 = 81$ | $p_A = 41/81$ | $p_B = 40/81$ | $\Delta p = 1/81$ ≈ | 1.2% |
| ξ = 5 | $3^5 = 3^5 = 243$ | $p_A = 122/243$ | $p_B = 121/243$ | $\Delta p = 1/243$ ≈ | 0.4% |
| ξ = 6 | $3^6 = 3^6 = 729$ | $p_A = 365/729$ | $p_B = 364/729$ | $\Delta p = 1/729$ ≈ | 0.1% |

NOTE – Simplifying, we assume the equal probabilities $p_{(J...J)} = p_{(J...K)} = p_{(K...J)} = 1/3$.

TABLE VI
SHORT WORDS

| Valid Images | Balance | Pattern w/L | Usage | J...J | J...K | K...J |
|---|---|---|---|---|---|---|
| K J K J K J | 3J-3K $^{inv}$ | L [−1] H | applicable | | | L [−1] H |
| K J K J J J | 4J-2K | L [=0] L | applicable | | | L [=0] L |
| K J J K J J | 4J-2K | L [−2] L | excluded | | | |
| K J J J K J | 4J-2K | L [=0] L | applicable | | | L [=0] L |
| K J J J J J | 5J-1K $^{inv}$ | L [−1] H | applicable | | | L [−1] H |
| J K J K J K | 3J-3K $^{inv}$ | L [+1] H | applicable | | L [+1] H | |
| J K J K J J | 4J-2K | L [=0] L | applicable | L [=0] L | | |
| J K J J K J | 4J-2K | L [+2] L | excluded | | | |
| J K J J J K | 4J-2K | L [=0] L | applicable | | L [=0] L | |
| J K J J J J | 5J-1K $^{inv}$ | L [−1] H | applicable | L [+1] H | | |
| J J K J K J | 4J-2K | L [=0] L | applicable | L [=0] L | | |
| J J K J J K | 4J-2K | L [−2] L | excluded | | | |
| J J K J J J | 5J-1K $^{inv}$ | L [−1] H | applicable | L [−1] H | | |
| J J J K J K | 4J-2K | L [=0] L | applicable | | L [=0] L | |
| J J J K J J | 5J-1K $^{inv}$ | L [+1] H | applicable | L [+1] H | | |
| J J J J K J | 5J-1K $^{inv}$ | L [−1] H | applicable | L [−1] H | | |
| J J J J J K | 5J-1K $^{inv}$ | L [+1] H | applicable | | L [+1] H | |
| J J J J J J | 6J-0K | L [=0] L | applicable | L [=0] L | | |

NOTE – Short words are six-letter = three-bit long, an avg. between RMII and MII.

TABLE VII
EXAMPLE LINGUISTIC MULTIPLEXING VARIANTS

| $m$ | $N_Q \leq N_E \leq N_A$ | $m_a + m_r$ | $n_e @ E = 2^1 \rightarrow n_D, k = m_r/n_r$ | | | $k \times n_D \cdot m \cdot BT$ |
|---|---|---|---|---|---|---|
| 2 | $5 = 2^0 \cdot 5^1$ | 0 + 2 | $n_e \geq 3.2$ | $n_D = 4$ | $k = 2$ | $2 \times 0.8 = 1.6$ μs |
| 3 | $11 = 2^0 \cdot 11^1$ | 0 + 3 | $n_e \geq 2.2$ | $n_D = 3$ | $k = 3$ | $3 \times 0.9 = 2.7$ μs |
| 3 | $10 = 2^1 \cdot 5^1$ | 1 + 2 | $n_e \geq 3.2$ | $n_D = 4$ | $k = 2$ | $2 \times 1.2 = 2.4$ μs |
| 3 | $9 = 2^0 \cdot 3^2$ | 0 + 3 | $n_e \geq 5.9$ | $n_D = 6$ | $k = 3$ | $3 \times 1.8 = 5.4$ μs |
| 4 | $27 = 2^0 \cdot 3^3$ | 0 + 4 | $n_e \geq 1.4$ | $n_D = 2$ | $k = 4$ | $4 \times 0.8 = 3.2$ μs |
| 4 | $25 = 2^0 \cdot 5^2$ | 0 + 2×2 | $n_e \geq 3.2$ | $n_D = 4$ | $k = 2$ | $2 \times 1.6 = 3.2$ μs |
| 4 | $24 = 2^3 \cdot 3^1$ | 3 + 1 | $n_e \geq 1.8$ | $n_D = 2$ | $k = 1$ | $1 \times 0.8 = 0.8$ μs |
| 4 | $22 = 2^1 \cdot 11^1$ | 1 + 3 | $n_e \geq 2.2$ | $n_D = 3$ | $k = 3$ | $3 \times 1.2 = 3.6$ μs |
| 4 | $21 = 2^0 \cdot 3^1 \cdot 7^1$ | 0 + 4 | $n_e \geq 2.6$ | $n_D = 3$ | $k = 4$ | $4 \times 1.2 = 4.8$ μs |
| 5 | $40 = 2^3 \cdot 5^1$ | 3 + 2 | $n_e \geq 3.2$ | $n_D = 4$ | $k = 2$ | $2 \times 2.0 = 4.0$ μs |
| 5 | $36 = 2^2 \cdot 3^2$ | 2 + 3 | $n_e \geq 5.9$ | $n_D = 6$ | $k = 3$ | $3 \times 3.0 = 9.0$ μs |
| 6 | $81 = 2^0 \cdot 3^4$ | 0 + 3×2 | $n_e \geq 5.9$ | $n_D = 6$ | $k = 3$ | $3 \times 3.6 = 10.8$ μs |
| 6 | $80 = 2^4 \cdot 5^1$ | 4 + 2 | $n_e \geq 3.2$ | $n_D = 4$ | $k = 2$ | $2 \times 2.4 = 4.8$ μs |
| 7 | $144 = 2^4 \cdot 3^2$ | 4 + 3 | $n_e \geq 5.9$ | $n_D = 6$ | $k = 3$ | $3 \times 4.2 = 12.6$ μs |
| 8 | $368 = 2^4 \cdot 23^1$ | 4 + 4 | $n_e \geq 2.0$ | $n_D = 2$ | $k = 4$ | $4 \times 1.6 = 6.4$ μs |
| 8 | $361 = 2^0 \cdot 19^2$ | 0 + 4×2 | $n_e \geq 4.1$ | $n_D = 5$ | $k = 4$ | $4 \times 4.0 = 16.0$ μs |
| 8 | $336 = 2^4 \cdot 3^1 \cdot 7^1$ | 4 + 4 | $n_e \geq 2.6$ | $n_D = 3$ | $k = 4$ | $4 \times 2.4 = 9.6$ μs |
| 8 | $324 = 2^2 \cdot 3^4$ | 2 + 3×2 | $n_e \geq 5.9$ | $n_D = 6$ | $k = 3$ | $3 \times 4.8 = 14.4$ μs |
| 8 | $320 = 2^6 \cdot 5^1$ | 6 + 2 | $n_e \geq 3.2$ | $n_D = 4$ | $k = 2$ | $2 \times 2.4 = 4.8$ μs |

NOTE – Further to the listed, other variants are also possible with $m$ = 4, 5, 6, 7, 8.

TABLE VIII
PCS MICROFRAMING GROUND

| $m$ | $N_C$ | $\log_2[\text{\# of }\langle J...\rangle]$ | $\log_2[\text{\# of }\langle J...J\rangle]$ | $\log_2[\text{\# of }\langle J...K\rangle]$ | $m \cdot BT$ |
|---|---|---|---|---|---|
| 9 | $2^9$ | ≈ 9.9 < $m$ + 1 | ≈ 9.2 < $m$ + 1 | ≈ 8.5 < $m$ + 1 | 0.9 μs |
| 10 | $2^{10}$ | ≈11.2 < $m$ + 1 | ≈10.5 < $m$ + 1 | ≈ 9.8 < $m$ + 1 | 1.0 μs |
| 11 | $2^{11}$ | ≈12.5 < $m$ + 1 | ≈11.8 < $m$ + 1 | ≈11.1 < $m$ + 1 | 1.1 μs |
| 12 | $2^{12}$ | ≈13.8 < $m$ + 1 | ≈13.1 > $m$ + 1 | ≈12.4 < $m$ + 1 | 1.2 μs |
| 13 | $2^{13}$ | ≈15.2 > $m$ + 2 | ≈14.5 > $m$ + 1 | ≈13.8 < $m$ + 1 | 1.3 μs |
| 14 | $2^{14}$ | ≈16.5 > $m$ + 2 | ≈15.8 > $m$ + 1 | ≈15.1 > $m$ + 1 | 1.4 μs |
| 15 | $2^{15}$ | ≈17.8 > $m$ + 2 | ≈17.1 > $m$ + 2 | ≈16.4 > $m$ + 1 | 1.5 μs |
| 16 | $2^{16}$ | ≈19.2 > $m$ + 3 | ≈18.5 > $m$ + 2 | ≈17.8 > $m$ + 1 | 1.6 μs |
| 17 | $2^{17}$ | ≈20.5 > $m$ + 3 | ≈19.8 > $m$ + 2 | ≈19.1 > $m$ + 2 | 1.7 μs |
| 18 | $2^{18}$ | ≈21.9 > $m$ + 3 | ≈21.2 > $m$ + 3 | ≈20.5 > $m$ + 2 | 1.8 μs |
| 19 | $2^{19}$ | ≈23.2 > $m$ + 4 | ≈22.5 > $m$ + 3 | ≈21.8 > $m$ + 2 | 1.9 μs |
| 20 | $2^{20}$ | ≈24.6 > $m$ + 4 | ≈23.9 > $m$ + 3 | ≈23.2 > $m$ + 3 | 2.0 μs |

NOTE – We count only balanced, i.e., those of pattern ..[=0].., valid serial images.

TABLE IX
NORMAL WORDS

| Mask | Pattern w/L | # of | Applicable Images | Balance | J . . . | . . . J |
|---|---|---|---|---|---|---|
| J...J (Σ 17) | L [=0] L | (7) | J K J K J J J J | 6J-2K | $A_1$ | $B_1$ |
| | | | J K J J K J J J | 6J-2K | $A_2$ | $B_2$ |
| | | | J K J J J K J J | 6J-2K | $A_3$ | $B_3$ |
| | | | J K J J J J K J | 6J-2K | $A_4$ | $B_4$ |
| | | | J J J K J K J J | 6J-2K | $A_5$ | $B_5$ |
| | | | J J J J K J K J | 6J-2K | $A_6$ | $B_6$ |
| | | | J J J J J J J J | 8J-0K | $A_7$ | $B_7$ |
| | | | J K J K J J K J | 5J-3K inv | $A_8$ | $B_8$ |
| | | | J J K J K J K J | 5J-3K inv | $A_9$ | $B_9$ |
| | | | J J K J J J J J | 7J-1K inv | $A_{10}$ | $B_{10}$ |
| | | | J J J K J J J J | 7J-1K inv | $A_{11}$ | $B_{11}$ |
| | L [−1] H | (5) | J J J J J K J J | 7J-1K inv | $A_{12}$ | $B_{12}$ |
| | L [+1] H | (5) | J K J K J K J J | 5J-3K inv | $A_{13}$ | $B_{13}$ |
| | | | J K J J K J K J | 5J-3K inv | $A_{14}$ | $B_{14}$ |
| | | | J K J J J J J J | 7J-1K inv | $A_{15}$ | $B_{15}$ |
| | | | J J J K J J J J | 7J-1K inv | $A_{16}$ | $B_{16}$ |
| | | | J J J J J K J J | 7J-1K inv | $A_{17}$ | $B_{17}$ |
| J...K (Σ 10) | L [=0] L | (4) | J K J K J K J K | 4J-4K | $A_{18}$ | |
| | | | J K J J J J J K | 6J-2K | $A_{19}$ | |
| | | | J J J K J J J K | 6J-2K | $A_{20}$ | |
| | | | J J J J J K J K | 6J-2K | $A_{21}$ | |
| | L [−1] H | (1) | J J J K J K J K | 5J-3K inv | $A_{22}$ | |
| | L [+1] H | (5) | J K J K J J J K | 5J-3K inv | $A_{23}$ | |
| | | | J K J J J K J K | 5J-3K inv | $A_{24}$ | |
| | | | J J K J J J J K | 5J-3K inv | $A_{25}$ | |
| | | | J J K J J J K K | 5J-3K inv | $A_{26}$ | |
| | | | J J J J J J J K | 7J-1K inv | $A_{27}$ | |
| K...J (Σ 10) | L [=0] L | (4) | K J K J K J K J | 4J-4K | | $B_{18}$ |
| | | | K J K J J J J J | 6J-2K | | $B_{19}$ |
| | | | K J J J K J J J | 6J-2K | | $B_{20}$ |
| | | | K J J J J J K J | 6J-2K | | $B_{21}$ |
| | | | K J K J K J J J | 5J-3K inv | | $B_{22}$ |
| | | | K J K J J J K J | 5J-3K inv | | $B_{23}$ |
| | | | K J J J K J K J | 5J-3K inv | | $B_{24}$ |
| | | | K J J J J K J J | 5J-3K inv | | $B_{25}$ |
| | L [−1] H | (5) | K J J J J J J J | 7J-1K inv | | $B_{26}$ |
| | L [+1] H | (1) | K J K J J J J J | 5J-3K inv | | $B_{27}$ |

NOTE – Normal words are eight-letter = nibble long, relate with the MII path width.
NOTE – Valid serial images with a DC bias outside [±1] are deliberately excluded.

TABLE XI
LONG WORDS

| Mask | Balance | # of | Applicable [Valid Serial] Images | A/B Index |
|---|---|---|---|---|
| J...J (Σ 233) | 10J-6K | (40) | J K J K J K J K J K J J J J | 1 / 1 |
| | | | . . . | . . . |
| | | | J J J J K J K J K J K J K J | 40 / 40 |
| | 12J-4K | (150) | J K J K J K J J J J J J J J | 41 / 41 |
| | | | . . . | . . . |
| | | | J J J J J J J K J K J K J K | 190 / 190 |
| | 14J-2K | (42) | J K J K J J J J J J J J J J | 191 / 191 |
| | | | . . . | . . . |
| | | | J J J J J J J J J J J K J K | 232 / 232 |
| | 16J-0K | (1) | J J J J J J J J J J J J J J | 233 / 233 |
| J...K (Σ 143) | 8J-8K | (1) | J K J K J K J K J K J K J K | 234 / — |
| | 10J-6K | (60) | J K J K J K J K J J J J J K | 235 / — |
| | | | . . . | . . . |
| | | | J J J J K J K J K J K J K | 294 / — |
| | 12J-4K | (75) | J K J K J K J J J J J J J K | 295 / — |
| | | | . . . | . . . |
| | | | J J J J J J J J K J K J K | 369 / — |
| | 14J-2K | (7) | J K J J J J J J J J J J J K | 370 / — |
| | | | . . . | . . . |
| | | | J J J J J J J J J J J K J K | 376 / — |
| K...J (Σ 143) | 8J-8K | (1) | K J K J K J K J K J K J K J | — / 234 |
| | 10J-6K | (60) | K J K J K J K J K J J J J J | — / 235 |
| | | | . . . | . . . |
| | | | K J J J J J K J K J K J K J | — / 294 |
| | 12J-4K | (75) | K J K J K J J J J J J J J J | — / 295 |
| | | | . . . | . . . |
| | | | K J J J J J J J J J K J K J | — / 369 |
| | 14J-2K | (7) | K J K J J J J J J J J J J J | — / 370 |
| | | | . . . | . . . |
| | | | K J J J J J J J J J J J K J | — / 376 |

TOTAL:

| J... or ...J (Σ 376) | 8J-8K | (1) | $< 2^8$ | |
| 10J-6K | (100) | $< 2^8$ | (Σ 325) > $2^8$ |
| 12J-6K | (225) | $< 2^8$ | |
| 14J-2K | (49) | $< 2^8$ | (Σ 278) > $2^8$ |
| 16J-0K | (1) | $< 2^8$ | |

(Σ 326) > $2^8$
$2^8 = 256$ is the data modulus, (Σ...) is the transport capacity

NOTE – Long words are sixteen-letter = octet long, relate with the MAC data unit.
NOTE – Valid serial images with a non-zero DC bias are all deliberately excluded.

TABLE X
EXAMPLE CODING STAGES BUILT AROUND NORMAL WORDS

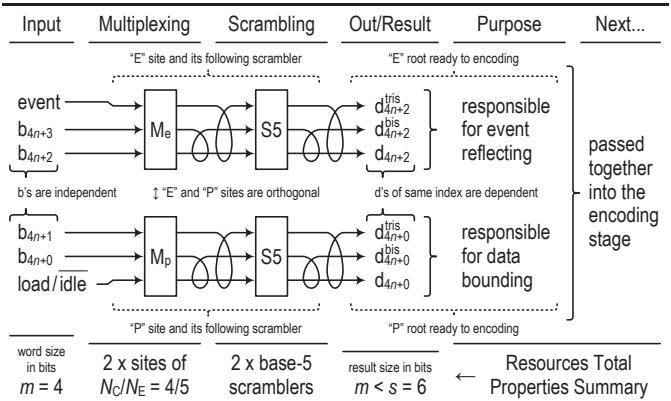

NOTE – There are $n_r = 2$ roots together conveying $m_r = 2 \times n_r = 4$ information bits.
NOTE – There are no bits assigned to the affix morpheme, i.e., $m_a = m - m_r = 0$.
NOTE – We assume $k = 2$ rounds and $E = 2^1$ echo modulus cancelled per a round.
NOTE – Required / selected (integer) round duration are $n_e \geq 3.2 / n_D = 4$ words.
NOTE – Echo multiplexing process is $1/2^2 \to 2^2 \cdot 2^4 / 2^1 \to 1/2^1 \to 2^1 \cdot 2^4 / 2^0 \to 1/2^0$.

TABLE XII
EXAMPLE CODING STAGES BUILT AROUND LONG WORDS

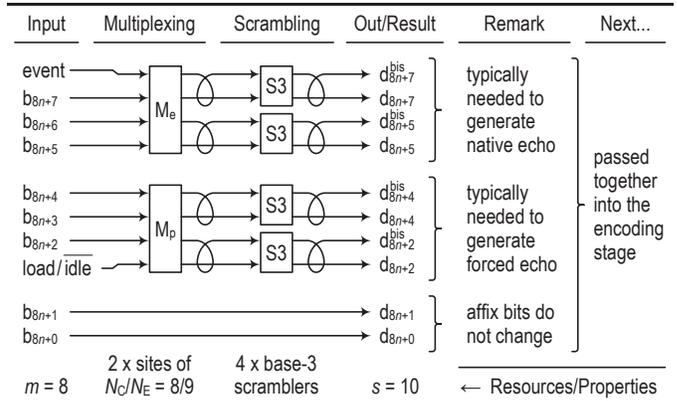

NOTE – There are $n_r = 2$ roots together conveying $m_r = 3 \times n_r = 6$ information bits.
NOTE – There are $m_a = 2$ bits assigned to the affix morpheme, $m_a + m_r = m = 8$.
NOTE – We assume $k = 3$ rounds and $E = 2^1$ echo modulus cancelled per a round.
NOTE – Required / selected (integer) round duration are $n_e \geq 5.9 / n_D = 6$ words.
NOTE – Echo multiplexing process is $1/2^3 \to 2^3 \cdot 2^{18} / 2^2 \to \ldots \to 2^1 \cdot 2^{18} / 2^0 \to 1/2^0$.

TABLE XIII
NUMERICAL COMPARISON

| Word Size, $m$, in independent information bits → | $m=1$ | $m=2$ | $m=3$ | $m=4$ | $m=5$ | $m=6$ | $m=7$ | $m=8$ | $m=9$ | $m=10$ |
|---|---|---|---|---|---|---|---|---|---|---|
| Data Modulus, $N_C = 2^m$ | 2 | 4 | 8 | 16 | 32 | 64 | 128 | 256 | 512 | 1,024 |
| Required Transport Capacity, $N_Q \geq N_C + 1$ | $\geq 3$ | $\geq 5$ | $\geq 9$ | $\geq 17$ | $\geq 33$ | $\geq 65$ | $\geq 129$ | $\geq 257$ | $\geq 513$ | $\geq 1{,}025$ |
| Serial Image Length, $M = 2m$, in letters | 2 | 4 | 6 | 8 | 10 | 12 | 14 | 16 | 18 | 20 |
| Word / Bit Time Period, in µs | 0.1/0.1 | 0.2/0.1 | 0.3/0.1 | 0.4/0.1 | 0.5/0.1 | 0.6/0.1 | 0.7/0.1 | 0.8/0.1 | 0.9/0.1 | 1.0/0.1 |
| Letter Time Period, in µs    transport alphabet = { J, K } | 0.05 | 0.05 | 0.05 | 0.05 | 0.05 | 0.05 | 0.05 | 0.05 | 0.05 | 0.05 |
| Number of Valid Serial Images, including: | 3 | 7 | 18 | 47 | 123 | 322 | 843 | 2,207 | 5,778 | 15,127 |
| • of mask ‹J . . . J›, Σ over: | Σ 1 | Σ 3 | Σ 8 | Σ 21 | Σ 55 | Σ 144 | Σ 377 | Σ 987 | Σ 2,584 | Σ 6,765 |
| • • of pattern . . [=0] . .    balanced | 1 | 1 | 3 | 7 | 16 | 39 | 95 | 233 | 577 > $N_Q$ | 1,436 |
| • • of pattern . . [±1] . .    unbalanced | 0 | 2 | 4 | 10 | 26 | 64 | 160 | 402 | 1,010 | 2,546 |
| • • of other unbalanced pattern(s)    unbalanced | 0 | 0 | 1 | 4 | 13 | 41 | 122 | 352 | 997 | 2,783 |
| • of mask ‹J . . . K› or ‹K . . . J›, Σ over: | Σ 1 | Σ 2 | Σ 5 | Σ 13 | Σ 34 | Σ 89 | Σ 233 | Σ 610 | Σ 1,597 | Σ 4,181 |
| • • of pattern . . [=0] . .    balanced | 0 | 1 | 2 | 4 | 10 | 24 | 58 | 143 | 354 | 881 |
| • • of pattern . . [±1] . .    unbalanced | 1 | 1 | 2 | 6 | 14 | 35 | 89 | 224 | 567 | 1,439 |
| • • of other unbalanced pattern(s)    unbalanced | 0 | 0 | 1 | 3 | 10 | 30 | 86 | 243 | 676 | 1,861 |
| • of PAGE mask ‹J . . . .› or ‹. . . . J›, Σ over: | Σ 2 | Σ 5 | Σ 13 | Σ 34 | Σ 89 | Σ 233 | Σ 610 | Σ 1,597 | Σ 4,181 | Σ 10,946 |
| • • of pattern . . [=0] . .    balanced | 1 | 2 | 5 | 11 | 26 | 63 | 153 > $N_Q$ | 376 | 931 | $N_C \cdot 2^1$ < 2,317 |
| • • of pattern . . [±1] . .    unbalanced | 1 | 3 | 6 | 16 | 40 | 99 | 249 | 626 | 1,577 | 3,985 |
| • • of other unbalanced pattern(s)    unbalanced | 0 | 0 | 2 | 7 | 23 | 71 | 208 | 595 | 1,673 | 4,644 |
| Number of Applicable Images, $N_A$, per a page | 2 | 5 | 11 | 27 | 46 | 162 | 153 | 376 | under studying | |
| Suitable Transport Capacity, $N_Q \leq N_E \leq N_A$, if any | absent | single | variants | variants | variants | variants | variants | variants | under studying | |

(annotations in the table: "1 < $N_Q$", "2 = $N_Q$", "5 > $N_Q$")

equal in absolute nominal amplitudes but opposite in relative directions designated to them.

Introducing a simple, "jump-and-keep" transport alphabet consisting of only two letters, J and K, we associate a narrow pulse with a JJ letter sequence and a wide pulse, respectively, with a JKJ sequence, definitely consuming exact one letter per a half of a bit time period—that corresponds to one letter per a narrow pulse versus two letters per a wide pulse—jointly by (a) going overlapped, or glued, with exact one letter, which is always J, and (b) shifted in phase by a quarter bit time period, compared with the original MDI pulse train, see Table II, and then conclude that it is both possible and sufficient to describe a 10BASE-T-compatible signaling in an abstract form of some continuous text written on the JK alphabet, barely preventing any run of two and more consecutive K's in such the text we also refer to as the letter transport stream.

Grounding on this concept, we now can design some like-a-Manchester (LaM), block-based coding means built around an appropriate transport dictionary that consists exclusively of the so called valid serial images, i.e., the permitted words of a regular length measured as the number of letters within such a word, which are picked out of all the possible words of the selected length, written on the JK alphabet, to match with the compatibility criterion we established above.

## POSSIBLE CODING SCHEMES

With the LaM approach enabled, there are many solutions becoming possible, see Table XIII, and the designer takes the choice among variants, balancing the implementation between its complexity and reliability, by (a) varying the length of and applicability (filtering) criteria for the valid serial images, see Tables III, VI, IX, and XI, that opts the data modulus and predefines the transport capacity available in such a condition, respectively, (b) picking between appropriate coding schemes and approaches like e.g. multiplexing, see Table IV, and (c) embedding and then tuning compensation measures like e.g. switched pages, see Table V, if it is necessary.

When the available transport capacity corresponding to the selected data modulus exceeds that modulus by at least a part of a bit, see Table VII, the designer reasonably could ground on the linguistic multiplexing model described in [3] and thus implement an appropriate coding means like the described in [4], that passes the data-related and event-related information sets implicitly intermixed via the echo cancellation procedure, see Tables III then IV, IX then X, and XI then XII.

When the available transport capacity exceeds the selected data modulus by at least one integral bit, see Table VIII, the designer could ground on a sort of PCS microframing like in 1000BASE-T1 or 10GBASE-T, considering every transmitted word a microframe, i.e., a time slot of a fixed, finite duration, smaller than the user frame, that conveys the data-related and event-related information sets explicitly separated.

## CONCLUSION

Thereby, in this paper we highlighted a way to extend the 10BASE-T MAU with an event transfer ability, based on the transport redundancy of Manchester line code as well as on the behavior specificity of that MAU. Also, the proposed way could be applicable to other protocols being originated from or similar to the mentioned MAU.

# Data Coding Means and Event Coding Means Multiplexed Over a 10BASE-T1S PMA-like Entity

Alexander Ivanov

*Abstract*—In this paper, we consider the 10BASE-T1S Ethernet physical layer, very redundant due to the 4B/5B over DME coding way the layer is based on, to extend it with a service new to it.

*Index Terms*—Ethernet, linguistic multiplexing, multiplexing, physical layer seamless preemption, preemption, physical layer time synchronization, synchronization, like-a-Manchester, LaM, line code balancing, balancing, 10BASE-T1S.

## INTRODUCTION

ORIGINAL-SPEED Ethernet, type 10BASE-T1S[1] [1] is a comparatively fresh physical layer capable to operate over a single unshielded twisted pair of conductors located in a harsh environment like of automotive or industrial.

Due to the 4B/5B over DME coding scheme employed by the layer, we face a very excessive redundancy during the data transmission phase in the stream, of exact half the dictionary, if we can count on all the words within, see Table I.

For making this the case, we imply we can involve a set of out-of-band signaling symbols,[2] in a manner like it is done in [2] or [3], freeing every in-band symbol, i.e., any word present in the dictionary, of its initial purpose, see Table II.

In the rest of the paper, we describe a way developing the original physical layer in its services and features, all over its original duty and objective, grounding on the design approach we consider in the series of the following steps.

## FROM CONVENTIONAL INTO PREEMPTABLE

Reexamining the original dictionary, see Table I again, we easily discern two non-intersecting and (therefore) orthogonal selections consisting of the same number, that is equal to half the dictionary, but of distinct words we gather guided by their common in the sense of their properties, see Table III.

Among the properties, as we think, the meaning of most of which is clear, there is just one, called the d.c. impact lasting time, that needs an extra explanation, see Table IV.

We connect one of the selections with the sub-stream which is responsible for transferring the preemptable payload in the multiplexed stream during the data transmission phase.

Recalling the fate of submission of many prior works to the peer reviewed journal, such a try with this one also promises no chance, probably.

Please sorry for the author has no time to find this work a new home, peer reviewed or not, except of arXiv, and just hopes their it meets its reader, one or maybe various, whom the author beforehand thanks for their regard.

A. Ivanov is with JSC Continuum, Yaroslavl, the Russian Federation.

Digital Object Identifier 10.48550/arXiv.yymm.nnnnn (bundle).

TABLE I
REFERENCE TRANSPORT DICTIONARY

| Clause 147 Definition | Serial Image FoW : LoW | J-K-$inv$ Balance | Impact $\Delta$dc-$inv$ | $\Delta$dc Peaks | Pattern ...[$\Delta$dc]... |
|---|---|---|---|---|---|
| ——— | J K J K J K J K | 5J-5K$_{inv}$ | +1 $_{inv}$ | +1 / — | L [+1] H |
| ——— | J K J K J J J K | 7J-3K$_{inv}$ | +1 $_{inv}$ | +1 / — | L [+1] H |
| ——— | J K J J J J K | 7J-3K$_{inv}$ | +1 $_{inv}$ | +1 / — | L [+1] H |
| ——— | J J J K J J J K | 7J-3K$_{inv}$ | +1 $_{inv}$ | +1 / — | L [+1] H |
| ——— | J J J J K J K | 7J-3K$_{inv}$ | +1 $_{inv}$ | +1 / — | L [+1] H |
| ESD ERR | J J J K J K J J | 7J-3K$_{inv}$ | +1 $_{inv}$ | +1 / — | L [+1] H |
| SYNC/COMMIT | J K J K J K J J J | 7J-3K$_{inv}$ | +1 $_{inv}$ | +1 / — | L [+1] H |
| Data 0001$_{(2)}$ | J J J K J K J J J | 7J-3K$_{inv}$ | +1 $_{inv}$ | +1 / — | L [+1] H |
| Data 0010$_{(2)}$ | J K J K J J K J | 7J-3K$_{inv}$ | +1 $_{inv}$ | +1 / — | L [+1] H |
| Data 0100$_{(2)}$ | J J J J K J J K | 7J-3K$_{inv}$ | +1 $_{inv}$ | +1 / — | L [+1] H |
| Data 1000$_{(2)}$ | J J J J J K J K | 7J-3K$_{inv}$ | +1 $_{inv}$ | +1 / — | L [+1] H |
| Data 0000$_{(2)}$ | J K J J J J J J | 9J-1K$_{inv}$ | +1 $_{inv}$ | +1 / — | L [+1] H |
| Data 0111$_{(2)}$ | J J J J J J J K | 9J-1K$_{inv}$ | +1 $_{inv}$ | +1 / — | L [+1] H |
| Data 1011$_{(2)}$ | J J J J J J K J | 9J-1K$_{inv}$ | +1 $_{inv}$ | +1 / — | L [+1] H |
| Data 1101$_{(2)}$ | J J J J K J J J | 9J-1K$_{inv}$ | +1 $_{inv}$ | +1 / — | L [+1] H |
| Data 1111$_{(2)}$ | J J K J J J J J | 9J-1K$_{inv}$ | +1 $_{inv}$ | +1 / — | L [+1] H |
| Data 0011$_{(2)}$ | J J J K J J K J | 8J-2K | =0 | +1 / — | L [=0] L |
| Data 0101$_{(2)}$ | J J J J K J J K | 8J-2K | =0 | +1 / — | L [=0] L |
| Data 0110$_{(2)}$ | J K J J J J J K | 8J-2K | =0 | +1 / — | L [=0] L |
| Data 1001$_{(2)}$ | J J J J K J K J | 8J-2K | =0 | +1 / — | L [=0] L |
| Data 1010$_{(2)}$ | J K J J J K J J | 8J-2K | =0 | +1 / — | L [=0] L |
| Data 1100$_{(2)}$ | J K J J J K J J | 8J-2K | =0 | +1 / — | L [=0] L |
| Data 1110$_{(2)}$ | J K J K J J J J | 8J-2K | =0 | +1 / — | L [=0] L |
| ESD OK/BRS | J J J J J J K J K | 8J-2K | =0 | +1 / — | L [=0] L |
| ESD JAB | J J J K J K J J J | 8J-2K | =0 | +1 / — | L [=0] L |
| ESD/HB | J J J J J J K J K | 8J-2K | =0 | +1 / — | L [=0] L |
| SSD | J K J K J J J K | 6J-4K | =0 | +1 / — | L [=0] L |
| BEACON | J K J K J K J J K | 6J-4K | =0 | +1 / — | L [=0] L |
| ——— | J K J J J K J K | 6J-4K | =0 | +1 / — | L [=0] L |
| ——— | J J J J K J K J K | 6J-4K | =0 | +1 / — | L [=0] L |
| ——— | J J J K J K J K | 6J-4K | =0 | +1 / — | L [=0] L |
| SILENCE | J J J J J J J J | 10J-0K | =0 | — / — | L [=0] L |

NOTE – Impact, peaks, and pattern of an image are given for the initial line state of L, for of H they are changing respectively.

Consequently, we connect the rest of the selections with the sub-stream which is responsible for transferring the preempting, express payload in the same multiplexed stream.

Because the selections, preemptable and express, are completely orthogonal to each other, we can find a set of allowable transits between them, easily, distinguishably, unambiguously, and then establish a stream preemption process we reasonably consider seamless, see Tables V and VI, respectively.

Based on the spoken above, we collect and then expose the requirements specifying an appropriate design we further will search for, to update the original one, see Table VII.

---

[1]Allows for 10 Mbps operation in the full duplex mode, in the half duplex mode using the original, CDMA/CD-supporting MAC, and in the half duplex mode using the newly introduced, advanced, PLCA-supporting MAC, enabled over two pairs, one pair, and one pair of media, respectively, see [1].

[2]In the scope of a given ME ( $f_0$ ) signal, including a differential option, as the considered in-band signaling procedure, meeting any ME ( $f_0/2$ ) symbol causes an encoding violation, but such a violation is very reliably detectable and distinguishable as out-of-band, and thus usable, see [2] and [3].

TABLE II
IMPLIED SIGNALING ROUTINE

| Phase → | Idle | Start-of-Stream signaling sequence | Management Commands | Payload (Data) | End-of-Stream signaling sequence | Idle |
|---|---|---|---|---|---|---|
| Encoding Type Modulation | 0Vdiff or Hi-Z | out-of-band DME ($f_0/2$) | in-band DME ($f_0$) | in-band DME ($f_0 = 1/\tau_0$) | out-of-band DME ($f_0/2$) | 0Vdiff or Hi-Z |
| NOTE – Mgmt. cmds are optional. | | | $\tau_0$ is payload letter time, s, thus $f_0$ is payload letter rate, Hz | | | |

TABLE III
COMPARISON BETWEEN SELECTIONS

| Word Selection | $p_J(L\cdot n + i), 0 \le i < L=10$ | ILT/$\tau_0$, min/avg/max | | | Aliases |
|---|---|---|---|---|---|
| All 32 valid words | 1..5 1..5 1..5 1..5 1..5 | — | 4.00 | ∞ | — |
| 16 of Cl. 147 Data | 1..5 1..7 1..6 1..6 1..7 | 2 | 5.16 | 18 | Ref(erence) |
| 16 of pattern L[+1]H | 1..5 1..5 1..5 1..5 1..5 | 2 | 4.57 | 18 | +[+]− , −[−]+ |
| 16 of pattern L[=0]L | 1..5 1..5 1..5 1..5 1..5 | — | 3.25 | 8 | +[=]+ , −[=]− |

TABLE IV
IMPACT LASTING TIME (ILT) MEANING

| Letter Sent $^{impact,\ if\ any}$ → | ... | J | J | K$^{+1}$ | J | ... | J | J | K$^{-1}$ | J | ... |
|---|---|---|---|---|---|---|---|---|---|---|---|
| Accumulated dc bias | — | — | — | +1 | +1 | ... | +1 | +1 | +1 | — | — |
| MDI signal waveform | | | | | | | | | | | |
| Time when dc bias ≠ 0 | | | $t_1$ | ← | ILT = $t_2 - t_1 \sim 2\tau_0$ | → | $t_2$ | | | | |

TABLE V
EXAMPLE TRANSITS

| Even $n$ ; Odd $n$ | ↔ | Even $n$ ; Odd $n$ | ↔ | Even $n$ ; Odd $n$ | ↔ | Any $n$ ⋯ Any $n$ |
|---|---|---|---|---|---|---|
| +[+]− ; −[−]+ | | +[+]− ; −[−]− | | −[=]− ; −[−]+ | | +[=]+ ⋯ +[=]+ |
| (self-balancing pairs) | | | | | | (self-balancing words) |
| −[−]+ ; +[+]− | | −[−]+ ; +[=]+ | | +[=]+ ; +[+]− | | −[=]− ⋯ −[=]− |

TABLE VI
EXAMPLE PREEMPTION PROCESS

| Sub-Stream | Pattern @ $n$ | Transit Into | Transit From | Comment |
|---|---|---|---|---|
| Preemptable | +[+]− ; −[−]+ | ($f_0/2$) → ($f_0$) | ($f_0$) → ($f_0/2$) | octet-balancing |
| Express | +[=]+ | −[−]+ → +[=]+ | +[=]+ → +[+]− | octet-KEEPING |
| Express, alt. | −[=]− | +[+]− → −[=]− | −[=]− → −[−]+ | octet-breaking |

TABLE VII
DESIGN SEARCH LIMITS

| Approach | Basis | OT | $L$ | $\tau_0$ | $f_0$ | avg ILT/$\tau_0$ max | Imp. | Peak | $\frac{20}{L}\log_2 N$ |
|---|---|---|---|---|---|---|---|---|---|
| Reference | DME | .8 | 10 | 40 | 25 | ~5  18 | +1 | +1 | 8 |
| Proposed | LaM | .8 | ≥2 | ≥40 | ≤25 | informative only | ±1 | ±1 | ≥8 |
| Measured in → | | μs | letters | ns | MHz | letter periods | unit dozes | unit levels | bits/OT |

NOTE – OT is the octet time, OT = 8·BT, where BT is the bit time, BT = 100 ns.

TABLE VIII
DME FRAMEWORK

↓Kernel then Detailed→

| | column=from | | | | FROM | Upward Rule | | | +[+]− , −[−]+ U Seeds D | | +[=]+ , −[=]− U Seeds D | |
|---|---|---|---|---|---|---|---|---|---|---|---|---|
| | | J$^e$ | J$^o$ | K | | → | J$^e$ = + | J$^o$ = + | K | | | |
| J$^e$ ← | | [·] | | | | J$^e$ = + | [·] | [·] | [·] | [·] | [·] | [·] |
| J$^o$ ← | | | [·] | [·] | | J$^o$ = + | [·] | [·] | [·] | | | |
| K ← | | | [·] | | | K = + | [·] | [·] | [·] | [·] | [·] | [·] |
| row=into | | | | | INTO | | | | | (0)f | (L−1)c | (0)f | (L−1)c |
| $L$ must be even | | | | | | F = C$^T$ | | | | | | |

TABLE IX
LaM ESSENTIALS

| col= from row= into | Kernel → | | Intermediate | | | | → Detailed |
|---|---|---|---|---|---|---|---|
| | J | K | $i$ ↓ $i+1$ | J$^e$ ↓ | J$^o$ ↓ | K$^e$ ↓ K$^o$ ↓ | |
| J ← | [{·}{·}] | | J$^e$ ← | [·] | [·] | | see LaM Framework |
| K ← | [{·}] | | J$^o$ ← | [·] | [·] | | |
| | | | K$^e$ ← | [·] | | | |
| | | | K$^o$ ← | [·] | | | |
| defines transit basics accounts nothing else | | | accounts impact polarity | | | | accounts impact polarity and magnitude |

TABLE X
LaM SPECIFICS

See Tables XI thru XXIV running the next page in the following order:

XI • → XXIV

## FROM MANCHESTER INTO LIKE-A-MANCHESTER

We replace the original line code option, that was of DME, with its advanced substitute, that will be of LaM [4], priorly considering the original dictionary a set of J–K-written, quasi base-21 words [5], see Table I again, and next describing the underlying framework compactly [6], see Table VIII.

Using the same description basics, we construct the substituting framework, now LaM-related, aimed to preserve all the sensible properties of the original coding, see Table VII again, as well as to expand over the original performance, palpably, see Table IX and briefly the tables mentioned in Table X.

Given a length expressed in letters, each of those should be either J or K [4], we construct a set of transport words, each of that length as well as of those properties, also considering the words quasi base-21 [5], based on the renewed framework we also manipulate compactly [6], see Table XI.

## FROM DETERMINISTIC INTO PROBABILISTIC

Applying the renewed framework with a word length in the scope of the search, we can receive up to four sets of such the words, distinct in their patterns and thence orthogonal in their contents, which are, however, so different in how they behave when sequenced, see Tables XII, XIII, and XIV.

Gathering over those unequal word sets, we obtain a couple of two supersets, or pages, also unequal because so purposed, one to maximize the effective size of the overall coding space while another to compensate during sequencing, then inscribe them into an appropriate or given scrambling space, balancing them via bubbling [7], see Tables XV, XVI, and XVII.

Mutually balanced, the two pages, maximizing and compensating, comprise the transport dictionary of the coding option, which we consider and tune, if and while necessary, matching between the corresponding per-$i$-th-letter-period probabilities, as close as it is possible, see Tables XVIII and XIX.

As the result of all the mentioned above, we reach a closely matched, highly balanced state of the design whose dictionary reflects a coding space variable in time, finitely and discretely, but probabilistically, see Tables XX, XXI, and XXII.

## FROM TRANSPARENT INTO MULTIPLEXED

The probabilistic side of the current, $n$-th-word-time-period transport capacity, inherent for such a design, leads us to use a sort of time division multiplexing, shaping the characteristics of a design as dependent on those of the words the dictionary of that design consists of, see Tables XXIII and XXIV.

TABLE XI
LaM Framework

TABLE XII
LaM Paths

| Mask→ | J⋯J | J⋯K | K⋯J | K⋯K |
|---|---|---|---|---|
| Cap. | $N_{JJ}$ > | $N_{JK}$ = | $N_{KJ}$ > | $N_{KK}$ |

TABLE XIII
Valid Transits Between Paths

| $n \setminus n{+}1$ | J⋯J | J⋯K | K⋯J | K⋯K |
|---|---|---|---|---|
| ⋯J | yes | yes | yes | yes |
| ⋯K | yes | yes | — | — |

TABLE XIV
Jump (J) Probability

| $i$ | J⋯J | J⋯K | K⋯J | K⋯K | All |
|---|---|---|---|---|---|
| 0 | $1_{exact}$ | $1_{exact}$ | — | $1_{exact}$ | .6 |
| 1 | .7 | .7 | $1_{exact}$ | — | .8 |
| 2 | .8 | .8 | .7 | .7 | .7 |
| 3 | .7 | .7 | .8 | .8 | .7 |

⋯⋯ NOTE – Fractional numbers are approximate. ⋯⋯

| $L{-}4$ | .7 | .8 | .7 | .8 | .7 |
| $L{-}3$ | .8 | .7 | .8 | .7 | .7 |
| $L{-}2$ | .7 | $1_{exact}$ | .7 | — | .8 |
| $L{-}1$ | $1_{exact}$ | — | $1_{exact}$ | $1_{exact}$ | .6 |

TABLE XV
LaM Pages

| Name→ | Page A = All | Page B |
|---|---|---|
| Paths | J⋯J, J⋯K, K⋯J, K⋯K | J⋯J, J⋯K |

TABLE XVI
Balancing Within Page B

| Transport Capacity | J⋯J | J⋯K |
|---|---|---|
| Effective (framework) | $N_{JJ}$ > | $N_{JK}$ |
| Nominal (scrambler) | $\frac{\varepsilon}{\varepsilon+\zeta}N_{scr}$ | $\frac{\zeta}{\varepsilon+\zeta}N_{scr}$ |

TABLE XVII
Balancing Within Page A

| Cap. | J⋯J | J⋯K | K⋯J | K⋯K |
|---|---|---|---|---|
| Eff. | $N_{JJ}$ > | $N_{JK}$ = | $N_{KJ}$ > | $N_{KK}$ |
| Nom. | $\frac{\alpha}{\sum \alpha+\delta}N_{scr}$ | $\frac{\beta}{\sum \alpha+\delta}N_{scr}$ | $\frac{\gamma}{\sum \alpha+\delta}N_{scr}$ | $\frac{\delta}{\sum \alpha+\delta}N_{scr}$ |

TABLE XVIII
LaM Dictionary

| Prob.→ | $p(A) = p(\cdots J) = p(\alpha \div \zeta)$ | $p(B)$ |
|---|---|---|
| Value | $\frac{\varepsilon \cdot (\alpha+\beta+\gamma+\delta)}{\varepsilon \cdot (\alpha+\beta+\gamma+\delta)+(\varepsilon+\zeta)\cdot(\beta+\delta)}$ | $1 - p(A)$ |

TABLE XIX
Balancing Within Dictionary

| Goal → | Solution |
|---|---|
| $p(J\cdots) = p(\cdots J)$ | $\zeta \cdot (\beta+\delta) = \varepsilon \cdot (\gamma-\beta)$ |
| $p_J(i) \approx const$ | varying $\alpha \div \zeta$ |

TABLE XXIII
LaM Performance, Bits/OT, Not Less Than

| $L$ | K⋯K | K⋯J | K⋯ | J⋯J | J⋯ | All | \multicolumn{4}{c}{Two Pages in A/B Ratio of} |
|---|---|---|---|---|---|---|---|---|---|---|
| | | | | | | | 5:5 | 6:4 | 7:3 | 8:2 |
| 2 | — | — | — | — | — | — | | | | |
| 3 | — | — | — | — | — | 6.6 | 3.9 | 4.5 | 5.1 | 5.6 |
| 4 | — | — | — | — | 5.0 | 7.9 | 6.6 | 6.8 | 7.1 | 7.4 |
| 5 | — | — | 4.0 | 4.0 | 6.3 | **9.2** | 8.0 | 8.2 | 8.5 | 8.8 |
| 6 | — | 3.3 | 3.3 | 5.2 | 7.7 | 9.3 | 8.6 | 8.7 | 8.9 | 9.0 |
| 7 | 2.8 | 2.8 | 5.7 | 6.6 | **8.0** | 9.8 | 9.0 | 9.2 | 9.4 | 9.5 |
| 8 | — | 5.0 | 5.8 | 7.0 | 8.6 | 10.0 | 9.3 | 9.5 | 9.6 | 9.7 |
| 9 | 4.4 | 5.1 | 7.0 | 7.6 | 8.8 | 10.3 | 9.6 | 9.8 | 9.8 | 10.0 |
| 10 | 3.1 | 6.3 | 7.1 | **8.0** | **9.2** | 10.4 | 9.9 | 10.0 | 10.1 | 10.2 |
| 11 | 5.4 | 6.5 | 7.8 | 8.4 | 9.4 | 10.6 | 10.0 | 10.2 | 10.3 | 10.4 |
| 12 | 5.0 | 7.2 | **8.0** | 8.6 | 9.7 | 10.6 | 10.2 | 10.3 | 10.4 | 10.5 |
| 13 | 6.2 | 7.3 | 8.4 | 8.9 | 9.8 | 10.8 | 10.3 | 10.4 | 10.5 | 10.6 |
| 14 | 6.1 | 7.8 | 8.6 | **9.1** | 10.0 | 10.8 | 10.4 | 10.5 | 10.6 | 10.7 |
| 15 | 6.9 | **8.0** | 8.8 | 9.3 | 10.1 | 10.9 | 10.5 | 10.6 | 10.7 | 10.8 |
| 16 | 6.9 | 8.3 | **9.0** | 9.5 | 10.2 | 11.0 | 10.6 | 10.7 | 10.8 | 10.8 |
| 17 | 7.4 | 8.5 | 9.2 | 9.6 | 10.3 | 11.0 | 10.7 | 10.8 | 10.8 | 10.9 |
| 18 | 7.5 | 8.7 | 9.3 | 9.7 | 10.4 | 11.1 | 10.8 | 10.8 | 10.9 | 10.9 |
| 19 | 7.9 | 8.8 | 9.5 | 9.9 | 10.5 | 11.1 | 10.8 | 10.9 | 10.9 | 11.0 |
| 20 | **8.0** | **9.0** | 9.6 | 10.0 | 10.6 | 11.1 | 10.9 | 10.9 | 11.0 | 11.0 |
| 40 | 10.0 | 10.4 | 10.7 | 10.9 | 11.2 | 11.5 | 11.4 | 11.4 | 11.4 | 11.5 |
| 80 | 10.9 | 11.2 | 11.3 | 11.4 | 11.6 | 11.7 | 11.6 | 11.7 | 11.7 | 11.7 |

NOTE – Ratio 5:5 numbers are asymptotic.

TABLE XXIV
ILT/$\tau_0$, Average

| K⋯K | K⋯J | K⋯ | J⋯J | J⋯ | All |
|---|---|---|---|---|---|
| — | — | — | — | — | — |
| 2.0 | — | 2.0 | — | — | 1.0 |
| — | 2.0 | 2.0 | — | 1.0 | 1.3 |
| 4.0 | 2.0 | 3.0 | 1.0 | 1.3 | **2.0** |
| — | 3.0 | 3.0 | 1.3 | 2.0 | 2.3 |
| 5.0 | 3.0 | 4.0 | 2.0 | **2.3** | 2.9 |
| 4.0 | 4.0 | 4.0 | 2.3 | **2.9** | 3.3 |
| 6.0 | 4.0 | 4.9 | 3.3 | 3.3 | 3.8 |
| 5.3 | 4.9 | 5.0 | **3.3** | **3.8** | 4.2 |
| 7.0 | 5.0 | 5.8 | 3.8 | 4.2 | 4.8 |
| 6.5 | 5.8 | **6.0** | 4.2 | 4.8 | 5.2 |
| 7.9 | 6.0 | 6.7 | 4.8 | 5.2 | 5.7 |
| 7.6 | 6.7 | 7.0 | **5.2** | 5.7 | 6.1 |
| 8.8 | **7.0** | 7.6 | 5.7 | 6.1 | 6.6 |
| 8.7 | 7.6 | **8.0** | 6.1 | 6.6 | 7.1 |
| 9.6 | 8.0 | 8.6 | 6.6 | 7.1 | 7.6 |
| 9.7 | 8.6 | 8.9 | 7.1 | 7.6 | 8.0 |
| 10.5 | 8.9 | 9.5 | 7.6 | 8.0 | 8.5 |
| **10.7** | **9.5** | 9.9 | 8.0 | 8.5 | 9.0 |
| 20.3 | 18.9 | 19.4 | 17.5 | 18.0 | 18.5 |
| 39.3 | 37.9 | 38.3 | 36.5 | 36.9 | 37.4 |

TABLE XXII
Transport Capacity

| Path | Max Eff. | Design-determined Effective Capacity = no. of distinct words | Page | Design-determined Effective |
|---|---|---|---|---|
| J⋯J | $N_{JJ} \geq 0$ | $0 \leq {}^A N_{JJ} \leq \min\{N_{JJ} ; {}^{\alpha}/_{\sum\alpha+\delta}N_{scr}\}$, $0 \leq {}^B N_{JJ} \leq \min\{N_{JJ} ; {}^{\varepsilon}/_{\varepsilon+\zeta}N_{scr}\}$ | A | ${}^A N = {}^A N_{JJ} + {}^A N_{JK} + {}^A N_{KJ} + {}^A N_{KK}$ |
| J⋯K | $N_{JK} \geq 0$ | $0 \leq {}^A N_{JK} \leq \min\{N_{JK} ; {}^{\beta}/_{\sum\alpha+\delta}N_{scr}\}$, $0 \leq {}^B N_{JK} \leq \min\{N_{JK} ; {}^{\zeta}/_{\varepsilon+\zeta}N_{scr}\}$ | B | ${}^B N = {}^B N_{JJ} + {}^B N_{JK}$ |
| K⋯J | $N_{KJ} \geq 0$ | $0 \leq {}^A N_{KJ} \leq \min\{N_{KJ} ; {}^{\gamma}/_{\sum\alpha+\delta}N_{scr}\}$ | | |
| K⋯K | $N_{KK} \geq 0$ | $0 \leq {}^A N_{KK} \leq \min\{N_{KK} ; {}^{\delta}/_{\sum\alpha+\delta}N_{scr}\}$ | Dictionary | ${}^D N = p(A)\cdot {}^A N + p(B)\cdot {}^B N$ |

TABLE XXI
Balance Search Methods

| Coarse | (Very) Fine |
|---|---|
| $\alpha, \beta, \gamma, \delta, \varepsilon, \zeta$, and $N_{scr}$ are integral numbers $\geq 0$ | $\alpha\cdot N_{scr}, \beta\cdot N_{scr}, \gamma\cdot N_{scr}, \delta\cdot N_{scr}, \varepsilon\cdot N_{scr}$, and $\zeta\cdot N_{scr}$ are integral numbers $\geq 0$ |

TABLE XX
Path Picking Proportion

| | $p(J\cdots J)$ : $p(J\cdots K)$ : $p(J\cdots K)$ : $p(K\cdots K)$ | when $p(J\cdots)=p(\cdots J)$ | $p(J\cdots J)$ when $p(J\cdots)\neq p(\cdots J)$ | $p(J\cdots K)$ when $p(J\cdots)\neq p(\cdots J)$ | $p(J\cdots K)$ when $p(J\cdots)\neq p(\cdots J)$ | $p(K\cdots K)$ when $p(J\cdots)\neq p(\cdots J)$ |
|---|---|---|---|---|---|---|
| do not rely on $\varepsilon$ or $\zeta$ | $\alpha+\beta+\delta$ : $\gamma$ : $\gamma$ : $\delta$  ;  $\alpha+\beta+2\gamma+2\delta$ | | $\frac{\varepsilon\cdot\alpha+\varepsilon\cdot(\beta+\delta)}{\varepsilon\cdot(\alpha+\beta+\gamma+\delta)+(\varepsilon+\zeta)\cdot(\beta+\delta)}$ | $\frac{\varepsilon\cdot\beta+\zeta\cdot(\beta+\delta)}{\varepsilon\cdot(\alpha+\beta+\gamma+\delta)+(\varepsilon+\zeta)\cdot(\beta+\delta)}$ | $\varepsilon\cdot\gamma$ | $\varepsilon\cdot\delta$ |

TABLE XXV
POSSIBLE CHANNELS

| Type | Bit Rate | Perf. | α | β | γ | δ | ε | ζ | Balanceables |
|------|----------|-------|---|---|---|---|---|---|--------------|
| CBR | constant | Basic | set | — | — | — | — | — | $p_J(1) \cdots p_J(L-2)$ |
| NOTE – When $\beta = \delta = 0$, | | Medium | set | set | — | — | — | — | $p_J(1) \cdots p_J(L-1)$ |
| $p(A) / p(B) = \varepsilon / (\varepsilon + \zeta)$, | | Medium | set | — | set | — | set | set | $p_J(0) \cdots p_J(L-1)$ |
| $\zeta \cdot (\alpha + \gamma) = \varepsilon \cdot \gamma, \ldots$ | VBR | variable | Medium | set | — | set | set | set | $p_J(0) \cdots p_J(L-1)$ |
| $\ldots p(J \cdots J) : p(J \cdots K) : p(K \cdots J)$ | | Medium | set | set | set | — | set | set | $p_J(0) \cdots p_J(L-1)$ |
| are $2\varepsilon - \zeta : \zeta : \zeta$ to $2\varepsilon + \zeta$. | | High | set | set | set | set | set | set | $p_J(0) \cdots p_J(L-1)$ |

TABLE XXVI
FUNCTIONALS CONVERTIBLE FROM PERFORMANCE

| Perf., bits/OT | Preembtable Sub-Stream | Express Sub-Stream | Multiplexing Procedure | Event Sub-Stream | Multiplexing Procedure |
|----|----|----|----|----|----|
| ≥ 8.0 | 8 payload bits | | | | |
| ≥ 8.1 | 8 payload bits | 8 payload bits | with echo | | |
| | 8 payload bits | | | implicit/spread | with echo |
| | 8 payload bits | 8 payload bits | with echo | implicit/spread | with echo |
| ≥ 9.0 | 8 load + 1 fec | | | | |
| | 8 payload bits | 8 payload bits | echo-free | | |
| | 8 payload bits | | | 1 dedicated bit | echo-free |
| ≥ 9.1 | 8 load + 1 fec | 8 load + 1 fec | with echo | | |
| | 8 payload bits | 8 payload bits | echo-free | implicit/spread | with echo |
| | 8 payload bits | 8 payload bits | with echo | 1 dedicated bit | echo-free |
| ≥ 9.6 | 8 payload bits | 8 load + 1 fec | echo-free | | |
| | 8 payload bits | 8 load + 1 fec | with echo | implicit/spread | with echo |
| | 8 payload bits | 8 load + 1 fec | echo-free | implicit/spread | with echo |
| ≥ 10.0 | 8 load + 1 fec | 8 load + 1 fec | echo-free | | |

TABLE XXVII
ADVANTAGES CONVERTIBLE FROM PERFORMANCE

| Perf. of LaM at $f_0 = f_{0,\text{ref}}$ | 8 | 8.5 | 9 | 9.5 | 10 | 10.5 | 11 | 11.5 | bits/OT |
|---|---|---|---|---|---|---|---|---|---|
| $f_0$ enough to perf. = ref. | 25 | ~24 | ~22 | ~21 | 20 | ~19 | ~18 | ~17 | MHz |
| Letter Rate = Bandwidth Reduction | (ref.) | −6% | −11% | −16% | −20% | −24% | −27% | −30% | — |

TABLE XXVIII
EXAMPLE LaM-BASED DESIGNS

| L | Type | α | β | γ | δ | ε | ζ | p(A) | $p_J(i)$ | Perf.≥ | ILT/$\tau_0$ |
|---|---|---|---|---|---|---|---|---|---|---|---|
| 5 | VBR | 5 | 1 | 2 | 2 | 3 | 1 | 5/7 | 5/7 exact | 8.5 | 1.7$_\text{avg}$ |
| 10 | CBR | 1 | — | 4 | — | 5 | 4 | .357 | .72 ± .02 | 9.2 | 4.2$_\text{avg}$ |
| 10 | VBR | 4 | 1 | 2 | 1 | 2 | 1 | .727 | .72 ± .03 | 10.1 | 4.0$_\text{avg}$ |
| 14 | CBR | 1 | — | 3 | — | 4 | 3 | .364 | .72 ± .03 | 10.0 | 6.0$_\text{avg}$ |
| 15 | VBR | 4 | 1 | 2 | 1 | 3 | 1 | .750 | .72 ± .04 | 10.7 | 6.4$_\text{avg}$ |
| 20 | CBR | 2 | — | 5 | — | 7 | 5 | .368 | .72 ± .03 | 10.6 | 8.8$_\text{avg}$ |
| 20 | VBR | 8 | 3 | 4 | 2 | 5 | 1 | .739 | .72 ± .03 | 11.0 | 8.8$_\text{avg}$ |

TABLE XXIX
LINGUISTIC MULTIPLEXING IN A VBR-BASED APPLICATION

| Scenario→ | Clear | Event | Echo Cancellation Round | | |
|---|---|---|---|---|---|
| Word Period | n | n | n + 1 | | n + $n_e$ |
| Page Selected | \$(n) | \$(n) | \$(n+1) | ⋯ | \$(n + $n_e$) |
| Word Set Used | \$data | \$data* | \$data** | | \$data** |
| Word Set Card. | \$$N_C$ | \$$N_R$ | \$$N_E$ | | \$$N_E$ |
| Echo Modulus | ⋯1⟶1⋯ | ⋯1⟶E⋯ | ⋯E⟶⟶E/ΔE⋯ | | |
| Echo Duration and other definitions… | $n_e$ : ΔE • $N_C(n+1)$ • … • $N_C(n + n_e)$ ≤ $N_E(n+1)$ • … • $N_E(n + n_e)$ | | $N_{\ldots}$ ≥ 1 | | E ≥ ΔE > 1 |
| | \$$N_C$ < \$$N_E$ ≤ \$$N_C$ + \$$N_R$ ≤ \$N | | | | |
| \$ = \$(n) = | $\begin{cases} A, \text{if the } (n-1)\text{-th word lasts with J} \\ B, \text{if the } (n-1)\text{-th word lasts with K} \end{cases}$ | | $N_{\ldots}(n) = \begin{cases} ^A N_{\ldots}(n), \text{if } \$(n) = A \\ ^B N_{\ldots}(n), \text{if } \$(n) = B \end{cases}$ | | |

NOTE – It is suitable to design a channel where \$$N_C$, E, and ΔE are powers of two.

TABLE XXX
ON-THE-FLY CAPACITY RECONCILIATION ALGORITHM

| Given | Curr. Stage | Input | Data Queue Content | Value Modulus | Indices Read | Write | Word Page | Output | Next Stage |
|---|---|---|---|---|---|---|---|---|---|
| pages available, \$N's ≥ 1: Σ\$N ≥ 3, and the efficiency required, 0 < η < 1 | INIT | | $B_q \leftarrow 0$ ; | $N_q \leftarrow 1$ ; | m ; | n ; | \$ | | Read |
| | Read | $B_\text{in}$ as 0 ≤ $B_\text{in}$ = $B_\text{in}(m) < N_\text{in} = N_\text{in}(m)$ | | | | | | | Enq. |
| | Enqueue | $B_q \leftarrow B_\text{in} \cdot N_q + B_q$ ; $N_q \leftarrow N_\text{in} \cdot N_q$ ; $m \leftarrow m+1$ ; ; | | | | | | | TEST |
| | TEST | $\begin{cases} \text{if } N_q < \eta \cdot \$N \cdot N_\#, \text{where } N_\# = \text{ceil}(N_q/\$N) \\ \text{otherwise} \end{cases}$ | | | | | | | Read Write |
| | Write | 0 ≤ $B_\text{out}$ = $B_\text{out}(n)$ = $B_q$ mod \$N < $N_\text{out}(n)$ = \$N as $B_\text{out}$ | | | | | | | Deq. |
| | Dequeue | $B_q \leftarrow B_q$ div \$N ; $N_q \leftarrow N_\#$ ;; $n \leftarrow n+1$ ; \$ ← \$(n) | | | | | | | TEST |

## CONCLUSION

So, we described a way, compatible with and operable over the 10BASE-T1S PMA sublayer as well as similar ME-based protocols, that enables a LaM-renewed data transfer channel, variable or constant in its bit rate,[3] see Table XXV.

Compared to the ME-based original, a LaM-based channel demonstrates a higher performance, whose extra part is usable in a manner backing a new service, an improved feature, or a combination thereof, see Tables XXVI and XXVII.

Thus, exactly the performance of a renewed channel in the first place among the sensitive requirements, including those of balancing, too, restricts the choice of an appropriate design providing such the channel, see Table XXVIII.

Implementing a suitable design, one could resolve a coding means based on the linguistic multiplexing approach, matured since [8], operating atop the corresponding channel directly or indirectly, see Tables XXIX and XXX, respectively.

[3]Primarily, we assume we design a VBR channel based upon the following probabilistic model opting the next page in the implicit favor of page A:

$$\begin{cases} p(A) = p(A) \times p(\cdots J/A) + p(B) \times p(\cdots J/B) \\ p(B) = p(A) \times p(\cdots K/A) + p(B) \times p(\cdots K/B) \end{cases},$$

however, when $\beta = \delta = 0$, we imply a CBR channel based upon the different model that, doubly oppositely, opts in the very explicit favor of page B:

$$\begin{cases} p(A) = p(A) \times p(\cdots K/A) + p(B) \times p(\cdots J/B) \\ p(B) = p(A) \times p(\cdots J/A) + p(B) \times p(\cdots K/B) \end{cases},$$

preserving all the rest principles we propose to balance a channel we design, via defining its respective properties, for both the models the same.

# Data Coding Means and Event Coding Means Multiplexed Over the 10BASE-T1L PMA Sublayer

Alexander Ivanov

*Abstract*—In the paper, we modify the 10BASE-T1L Ethernet physical layer redundant, due to the 4B3T encoding the layer is based on, sufficiently to extend it with a new service.

*Index Terms*—Ethernet, linguistic multiplexing, multiplexing, physical layer seamy preemption, preemption, physical layer time synchronization, synchronization, 10BASE-T1L.

## Introduction

ORIGINAL-SPEED Ethernet, type 10BASE-T1L[1] [1] is a long-reach (∼1000 m) physical layer capable to operate over a fieldbus-gauged twisted pair of conductors located in a harsh environment like of automotive or industrial.

The complete 10BASE-T1L Physical Layer is built around those two vertically stacked sublayers, 10BASE-T1L PCS and 10BASE-T1L PMA, assigned its physical coding and physical medium attachment sublayers, correspondingly.

That 10BASE-T1L PCS employs the 4B3T coding scheme to convert in between the stream of nibbles of bits, present on the PHY service interface, and the stream of triplets of ternary symbols, present on the PMA service interface.

That 10BASE-T1L PMA sublayer throws for every ternary symbol with a PAM-3 signaling level, thus, we say we practice a three-letter-size transport alphabet to next record on it a set of three-letter-long transport words, consequentially.

To keep the physical signal in the line dc-balanced, that set, or the whole transport dictionary, is divided into four distinct but intersecting subsets, or pages, as the ground for the means bounding the current running disparity of the stream.

In this paper, we discuss a bunch of ways intended to reach a multiservice layer design, whose underlying coding means will be compatible with the 10BASE-T1L initial duty, as well as operable over the 10BASE-T1L PMA sublayer.

## The Ground Behind Any Possible Way

Given with the disparity-addressed behavior of the original coding means, as it was mentioned earlier, and further based on the linguistic multiplexing approach, as it was developed up to in [2] and [3], we refine the transport dictionary, during that, consider every its subset a page, either small or large, of either 16 or 18 words, respectively, see Table I.

Recalling the fate of submission of many prior works to the peer reviewed journal, such a try with this one also promises no chance, probably.

Please sorry for the author has no time to find this work a new home, peer reviewed or not, except of arXiv, and just hopes there it meets its reader, one or maybe various, whom the author beforehand thanks for their regard.

A. Ivanov is with JSC Continuum, Yaroslavl, the Russian Federation.

Digital Object Identifier 10.48550/arXiv.yymm.nnnnn (bundle).

[1]Allows for 10 Mbps operation in the (true) full duplex mode, see [1].

TABLE I
MULTIPAGED TRANSPORT DICTIONARY

| P. | Set | Transport Words in $^{(P)}$M | $^{(P)}N_C$ | $^{(P)}N_R$ | $^{(P)}N_E$ | GCD | Applicable |
|---|---|---|---|---|---|---|---|
| 1 | $^{(1)}$M | $^{(1)}m_0 \cdots {}^{(1)}m_{15}$ | 16 | — | — | — | @ Σdc = 1 |
| 2 | $^{(2)}$M | $^{(2)}m_0 \cdots {}^{(2)}m_{15}, {}^{(2)}m_{16}, {}^{(2)}m_{17}$ | 16 + 2 | = | 18 | $2^1$ | @ Σdc = 2 |
| 3 | $^{(3)}$M | $^{(3)}m_0 \cdots {}^{(3)}m_{15}, {}^{(3)}m_{16}, {}^{(3)}m_{17}$ | 16 + 2 | = | 18 | $2^1$ | @ Σdc = 3 |
| 4 | $^{(4)}$M | $^{(4)}m_0 \cdots {}^{(4)}m_{15}$ | 16 | — | — | — | @ Σdc = 4 |

TABLE II
CURRENT PAGE CONTENT

| Page | Context | Scope | Accessible Words | Multiplexing | Ev. Flags |
|---|---|---|---|---|---|
| P($n$) | Σdc = P($n$) | $^{(P)}$M $\begin{cases} m_0 \cdots m_{15}, m_{16}, m_{17} \\ \text{or} \\ m_0 \cdots m_{15} \end{cases}$ | | POSSIBLE | up to two |
| | | | | impossible | — |

TABLE III
CODING SPACES

| Space | Size = Σ Reps. of m's | Reps. of $m_0$ | Reps. of $m_1$ | Reps. of $m_2$ | $\cdots$ |
|---|---|---|---|---|---|
| Plain | 9, 12, 16, or 18 | up to once | up to once | up to once | $\cdots$ |
| Cipher | always $2^5$ = 32 | $R(m_0)$ | $R(m_1)$ | $R(m_2)$ | $\cdots$ |

TABLE IV
STREAM SCRAMBLING SCHEME

| Space | Randomity Source | Bits Used | Bit Aliases | Scrambling Scope |
|---|---|---|---|---|
| Cipher | 10BASE-T1L PRNG | $Sy_n$[0:4] | $s_n$[0] to $s_n$[4] | Regular Data Only |

TABLE V
TIME MEMO

| PHY Protocol | MII Bit Time | Transport Letter Time | Word (Nibble) Time |
|---|---|---|---|
| 10BASE-T1L | BT = 100 ns | $1/f_0$ = ⅘·BT = 133⅓ ns | $T_w$ = 4·BT = 400 ns |

In the scope of the currently processed page, be it small or large, we operate on its words in a uniform manner, so being aware only of its size, i.e., the number of distinct words within that page, not of their certain definition, see Table II.[2]

Because the current page varies in time probabilistically, so does its size, too, necessitating us to design a coding measure capable to operate on a transmission channel characterized by a probabilistically variable word (bit) rate.

[2]Denoted by $n$ (or $t$) is the current word (or letter) time period index.

TABLE VI
REFERENCE TRANSPORT DICTIONARY
VALID DURING REGULAR CLAUSE 146 DATA STREAMING PHASE

| Image | $\Delta dc$ | $\Delta dc$ Pks. | Trs. | @ $\Sigma dc=1$ | @ $\Sigma dc=2$ | @ $\Sigma dc=3$ | @ $\Sigma dc=4$ |
|---|---|---|---|---|---|---|---|
| L L L | −3 | — / −3 | — | | | | $1001_{(2)}$ |
| L L z | −2 | — / −2 | 1 | | | | $0011_{(2)}$ |
| L z L | −2 | — / −2 | 2 | | | | $1101_{(2)}$ |
| z L L | −2 | — / −2 | 1 | | | | $1000_{(2)}$ |
| L L H | −1 | — / −2 | 1 | | | $0110_{(2)}$ | $0110_{(2)}$ |
| L z z | −1 | — / −1 | 1 | | $0101_{(2)}$ | $0101_{(2)}$ | $0101_{(2)}$ |
| L H L | −1 | — / −1 | 2 | | $1100_{(2)}$ | $1100_{(2)}$ | $1100_{(2)}$ |
| z L z | −1 | — / −1 | 2 | | $0000_{(2)}$ | $0000_{(2)}$ | $0000_{(2)}$ |
| z z L | −1 | — / −1 | 1 | | $1111_{(2)}$ | $1111_{(2)}$ | $1111_{(2)}$ |
| H L L | −1 | +1 / −1 | 1 | | | $1010_{(2)}$ | $1010_{(2)}$ |
| L z H | =0 | — / −1 | 2 | $0111_{(2)}$ | $0111_{(2)}$ | $0111_{(2)}$ | $0111_{(2)}$ |
| L H z | =0 | — / −1 | 2 | $0100_{(2)}$ | $0100_{(2)}$ | $0100_{(2)}$ | $0100_{(2)}$ |
| z L H | =0 | — / −1 | 2 | $0001_{(2)}$ | $0001_{(2)}$ | $0001_{(2)}$ | $0001_{(2)}$ |
| z z z | =0 | — / — | — | | | | |
| z H L | =0 | +1 / — | 2 | $1110_{(2)}$ | $1110_{(2)}$ | $1110_{(2)}$ | $1110_{(2)}$ |
| H L z | =0 | +1 / — | 2 | $0010_{(2)}$ | $0010_{(2)}$ | $0010_{(2)}$ | $0010_{(2)}$ |
| H z L | =0 | +1 / — | 2 | $1011_{(2)}$ | $1011_{(2)}$ | $1011_{(2)}$ | $1011_{(2)}$ |
| L H H | +1 | +1 / −1 | 1 | $0110_{(2)}$ | $0110_{(2)}$ | | |
| z z H | +1 | +1 / — | 1 | $0011_{(2)}$ | $0011_{(2)}$ | $0011_{(2)}$ | |
| z H z | +1 | +1 / — | 2 | $1101_{(2)}$ | $1101_{(2)}$ | $1101_{(2)}$ | |
| H L H | +1 | +1 / — | 2 | $1001_{(2)}$ | $1001_{(2)}$ | $1001_{(2)}$ | |
| H z z | +1 | +1 / — | 1 | $1000_{(2)}$ | $1000_{(2)}$ | $1000_{(2)}$ | |
| H H L | +1 | +2 / — | 1 | $1010_{(2)}$ | $1010_{(2)}$ | | |
| z H H | +2 | +2 / — | 1 | $0101_{(2)}$ | | | |
| H z H | +2 | +2 / — | 2 | $0000_{(2)}$ | | | |
| H H z | +2 | +2 / — | 1 | $1111_{(2)}$ | | | |
| H H H | +3 | +3 / — | — | $1100_{(2)}$ | | | |

TABLE VIII
BROADENED TRANSPORT DICTIONARY
VALID DURING REGULAR MULTISERVICE DATA STREAMING PHASE

| $\Delta dc$ | …=1 | $R_n$ | $R_{n-1}$ | …=2 | $R_n$ | $R_{n-1}$ | …=3 | $R_n$ | $R_{n-1}$ | …=4 | $R_n$ | $R_{n-1}$ |
|---|---|---|---|---|---|---|---|---|---|---|---|---|
| −3 | | | | | | | | | | $m_0$ | 2 | — |
| −2 | | | | | | | | | | $m_1$ | 2 | 5 |
| −2 | | | | | | | | | | $m_2$ | 2 | 4 |
| −2 | | | | | | | | | | $m_3$ | 2 | 5 |
| −1 | | | | $m_{17}$ | 1 | — | $m_0$ | 2 | 2 | $m_4$ | 2 | 3 |
| −1 | | | | $m_{16}$ | 2 | — | $m_1$ | 2 | 2 | $m_5$ | 2 | 3 |
| −1 | | | | $m_{15}$ | 1 | — | $m_2$ | 2 | 3 | $m_6$ | 2 | 3 |
| −1 | | | | $m_{14}$ | 1 | — | $m_3$ | 2 | 3 | $m_7$ | 2 | 3 |
| −1 | | | | $m_{13}$ | 2 | — | $m_4$ | 2 | 2 | $m_8$ | 2 | 3 |
| −1 | | | | $m_{12}$ | 1 | — | $m_5$ | 2 | 2 | $m_9$ | 2 | 3 |
| =0 | $m_{15}$ | 2 | — | $m_{11}$ | 2 | 3 | $m_6$ | 2 | 3 | $m_{10}$ | 2 | — |
| =0 | $m_{14}$ | 2 | — | $m_{10}$ | 2 | 3 | $m_7$ | 2 | 3 | $m_{11}$ | 2 | — |
| =0 | $m_{13}$ | 2 | — | $m_9$ | 2 | 3 | $m_8$ | 2 | 3 | $m_{12}$ | 2 | — |
| =0 | | | | | | | | | | | | |
| =0 | $m_{12}$ | 2 | — | $m_8$ | 2 | 3 | $m_9$ | 2 | 3 | $m_{13}$ | 2 | — |
| =0 | $m_{11}$ | 2 | — | $m_7$ | 2 | 3 | $m_{10}$ | 2 | 3 | $m_{14}$ | 2 | — |
| =0 | $m_{10}$ | 2 | — | $m_6$ | 2 | 3 | $m_{11}$ | 2 | 3 | $m_{15}$ | 2 | — |
| +1 | $m_9$ | 2 | 3 | $m_5$ | 2 | 2 | $m_{12}$ | 1 | — | | | |
| +1 | $m_8$ | 2 | 3 | $m_4$ | 2 | 2 | $m_{13}$ | 2 | — | | | |
| +1 | $m_7$ | 2 | 3 | $m_3$ | 2 | 3 | $m_{14}$ | 1 | — | | | |
| +1 | $m_6$ | 2 | 3 | $m_2$ | 2 | 3 | $m_{15}$ | 1 | — | | | |
| +1 | $m_5$ | 2 | 3 | $m_1$ | 2 | 2 | $m_{16}$ | 2 | — | | | |
| +1 | $m_4$ | 2 | 3 | $m_0$ | 2 | 2 | $m_{17}$ | 1 | — | | | |
| +2 | $m_3$ | 2 | 5 | | | | | | | | | |
| +2 | $m_2$ | 2 | 4 | | | | | | | | | |
| +2 | $m_1$ | 2 | 5 | | | | | | | | | |
| +3 | $m_0$ | 2 | — | | | | | | | | | |

TABLE VII
ORIGINAL CODING PORTRAIT
VALID DURING REGULAR CLAUSE 146 DATA STREAMING PHASE

| Probability | $t=3n+0$ | $t=3n+1$ | $t=3n+2$ | Average | Statistics |
|---|---|---|---|---|---|
| $p(\Sigma dc=0)$ | $^{12}/_{416}\approx .03$ | $^{8}/_{416}\approx .02$ | — | $^{20}/_{1248}\approx .02$ | $\mu_0=2.5$ |
| $p(\Sigma dc=1)$ | $^{65}/_{416}\approx .16$ | $^{65}/_{416}\approx .16$ | $^{4}/_{26}\approx .15$ | $^{194}/_{1248}\approx .16$ | $\Delta=\pm2.5$ |
| $p(\Sigma dc=2)$ | $^{131}/_{416}\approx .31$ | $^{135}/_{416}\approx .32$ | $^{9}/_{26}\approx .35$ | $^{410}/_{1248}\approx .33$ | |
| $p(\Sigma dc=3)$ | $^{131}/_{416}\approx .31$ | $^{135}/_{416}\approx .32$ | $^{9}/_{26}\approx .35$ | $^{410}/_{1248}\approx .33$ | $1\sigma\approx1.0$ |
| $p(\Sigma dc=4)$ | $^{65}/_{416}\approx .16$ | $^{65}/_{416}\approx .16$ | $^{4}/_{26}\approx .15$ | $^{194}/_{1248}\approx .16$ | $2\sigma\approx2.1$ |
| $p(\Sigma dc=5)$ | $^{12}/_{416}\approx .03$ | $^{8}/_{416}\approx .02$ | — | $^{20}/_{1248}\approx .02$ | $3\sigma\approx3.1$ |
| $p(\text{letter}=L)$ | $^{134}/_{416}\approx .32$ | $^{134}/_{416}\approx .32$ | $^{134}/_{416}\approx .32$ | $^{402}/_{1248}\approx .32$ | |
| $p(\text{letter}=z)$ | $^{148}/_{416}\approx .36$ | $^{148}/_{416}\approx .36$ | $^{148}/_{416}\approx .36$ | $^{444}/_{1248}\approx .36$ | letter-to-word |
| $p(\text{letter}=H)$ | $^{134}/_{416}\approx .32$ | $^{134}/_{416}\approx .32$ | $^{134}/_{416}\approx .32$ | $^{402}/_{1248}\approx .32$ | INDEX RATIO |
| $p(\text{transit after})$ | $\approx .79$ | $\approx .79$ | $\approx .68$ | $\approx .76$ | $\frac{t}{n}=\frac{3}{1}$ |
| | boundary | inter-letter | inter-letter | inter-word | |
| $p(\text{run of }C\text{ consecutive L's})$ | | present for $C\leq 5$, absent for $C>5$ | | | |
| $p(\text{run of }C\text{ consecutive z's})$ | | present for $C\leq 4$, absent for $C>4$ | | | |
| $p(\text{run of }C\text{ consecutive H's})$ | | present for $C\leq 5$, absent for $C>5$ | | | Remark |

TABLE IX
MODIFIED CODING PORTRAIT
VALID DURING REGULAR MULTISERVICE DATA STREAMING PHASE

| Probability | $t=3n+0$ | $t=3n+1$ | $t=3n+2$ | Average | Remark |
|---|---|---|---|---|---|
| $p(\Sigma dc=0)$ | $^{24}/_{832}\approx .03$ | $^{25}/_{832}\approx .03$ | — | $^{49}/_{2496}\approx .02$ | counted over cases of $R_n$'s only with cases of $R_{n-1}$'s not counted because as rare as negligible |
| $p(\Sigma dc=1)$ | $^{130}/_{832}\approx .16$ | $^{121}/_{832}\approx .15$ | $^{4}/_{26}\approx .15$ | $^{379}/_{2496}\approx .15$ | |
| $p(\Sigma dc=2)$ | $^{262}/_{832}\approx .31$ | $^{270}/_{832}\approx .32$ | $^{9}/_{26}\approx .35$ | $^{820}/_{2496}\approx .33$ | |
| $p(\Sigma dc=3)$ | $^{262}/_{832}\approx .31$ | $^{270}/_{832}\approx .32$ | $^{9}/_{26}\approx .35$ | $^{820}/_{2496}\approx .33$ | |
| $p(\Sigma dc=4)$ | $^{130}/_{832}\approx .16$ | $^{121}/_{832}\approx .15$ | $^{4}/_{26}\approx .15$ | $^{379}/_{2496}\approx .15$ | |
| $p(\Sigma dc=5)$ | $^{24}/_{832}\approx .03$ | $^{25}/_{832}\approx .03$ | — | $^{49}/_{2496}\approx .02$ | |
| $p(\text{letter}=L)$ | $^{277}/_{832}\approx .33$ | $^{268}/_{832}\approx .32$ | $^{277}/_{832}\approx .33$ | $^{822}/_{2496}\approx .33$ | |
| $p(\text{letter}=z)$ | $^{278}/_{832}\approx .33$ | $^{296}/_{832}\approx .36$ | $^{278}/_{832}\approx .33$ | $^{852}/_{2496}\approx .34$ | |
| $p(\text{letter}=H)$ | $^{277}/_{832}\approx .33$ | $^{268}/_{832}\approx .32$ | $^{277}/_{832}\approx .33$ | $^{822}/_{2496}\approx .33$ | |
| $p(\text{transit after})$ | $\approx .77$ | $\approx .77$ | $\approx .58$ | $\approx .71$ | |
| $p(\text{run of}\ldots\text{L's}), p(\text{run of}\ldots\text{z's}), p(\text{run of}\ldots\text{H's})$ | | present/absent for the same $C$'s | | | |

Considering since now for further the regular transmission only,[3] we define all the necessary coding spaces, see Table III, some plain of different sizes and some cipher of the same size equal to the single size of the scrambling space, see Table IV, assuming here that a cipher space, characterized by the single, nominal size, is a mapping of some corresponding plain space, characterized by its effective size, being bubbled (up) to match with the size of the former, like it is done in [4], i.e., between the base-32 space of the scrambler, seen in time, see Table V, and an any-base coding space, like it is done in [5].

Thus, the definition of the spaces stands behind the certain transport dictionary, inherited from and then enlarged over the original 10BASE-T1L one, we use to establish a multiservice coding means, along other, preserving the disparity-addressed behavior of the progenitor, see Table VIII first in conjunction and then in comparison with Table VI,[4] respectively.

Among a lot of variants of such the definition, possible and suitable, we select out just one, arbitrarily to some extent and therefore subjectively, which itself is generic, transparent, and clear, all in the scope of our discussion, of course, as well as, the same time, whose properties are very close to the original, see Table IX in comparison with Table VII.

---

[3]Regular data quantities are of those in modes SEND_N and SEND_I.

[4]Data binary values and their symbols are as they are after scrambling.

TABLE X
REFERENCE TRANSPORT DICTIONARY
VALID DURING REGULAR CLAUSE 146 DATA DELIMITING PHASE

| $s_n[4]$ | Prd. | Cond. | ... = 1 | Δdc | ... = 2 | Δdc | ... = 3 | Δdc | ... = 4 | Δdc |
|---|---|---|---|---|---|---|---|---|---|---|
| any | 1st | any | z z z | =0 | z z z | =0 | z z z | =0 | z z z | =0 |
| any | 2nd | any | z z z | =0 | z z z | =0 | z z z | =0 | z z z | =0 |
| 0 | 3rd | any | L z H | =0 | L z z | −1 | L z L | −2 | L L L | −3 |
|  | 4th | SSD | H H L | +1 |  |  |  |  |  |  |
|  |  | ESD | H L H | +1 |  |  |  |  |  |  |
|  |  | ESD_ERR | L H H | +1 |  |  |  |  |  |  |
| 1 | 3rd | any | H H H | +3 | H L H | +1 | H z z | +1 | H z L | =0 |
|  | 4th | SSD |  |  |  |  |  |  | L L H | −1 |
|  |  | ESD |  |  |  |  |  |  | L H L | −1 |
|  |  | ESD_ERR |  |  |  |  |  |  | H L L | −1 |

TABLE XI
HOPEFUL ECHO CANCELLATION ROUTINE
ONE SHOULD KNOW ABOUT HOWEVER PROBABLY WILL NEVER USE

| Make a try to obtain a result valid in the stream via pre-calculating the PRNG output for periods from $n+0$ to $n+5$ and pre-fetching the incoming data nibbles of the same periods | Successful? | Otherwise |
|---|---|---|
| Effective Echo Content: $\frac{1}{E} \to E/1 \cdot 2 \cdot (2 \cdot 2 \cdot 2 \cdot 2)^6 \cdot 1/E \to \frac{1}{E/2}$ arithmetic representation; $6 = \log_2 E < \infty$; $n_e \times k = n_D$; 18 $n+0$, 18 $n+1$, 18 $n+2$, 18 $n+3$, 18 $n+4$, 18 $n+5$ | Transmit the result of that try then $n \leftarrow n+6$ | Perform a mocking echo round (see below) then $n \leftarrow n+1$ |

TABLE XII
MOCKING ECHO CANCELLATION ROUTINE
COMPATIBLE WITH MULTISERVICE OPERATION

| $n_e \times k = n_D$ | Input Echo | One Round Depiction | Output Echo | Functional Equivalent |
|---|---|---|---|---|
| 1 | ∞ | ∞ | $1/E \to E/1 \cdot 2 \cdot 2 \cdot 2 \cdot 2 \cdot 1/E \to 1/E$ | Delay by $\log_2 E$ bit times |

TABLE XIII
ECHO CANCELLATION AREA
AVAILABLE DURING MULTISERVICE OPERATION

| Room Potentially Feasible → | Pr.+SFD of This Frame | IFG Next to This Frame |
|---|---|---|
| Room Size ( minus Cl. 146 delim. ) | up to 64 (48) bit times | up to 96 (80) bit times |

TABLE XIV
BROADENED TRANSPORT DICTIONARY
VALID DURING REGULAR MULTISERVICE EVENT INSERTION PHASE

| Space | ... = 1 | Δdc | ... = 2 | Δdc | ... = 3 | Δdc | ... = 4 | Δdc | Assembly |
|---|---|---|---|---|---|---|---|---|---|
| $R_{n-1}$ | $m_1 \div m_9$ | +1/+2 | $m_0 \div m_{11}$ | 0/+1 | $m_0 \div m_{11}$ | 0/−1 | $m_1 \div m_9$ | −1/−2 | fade-in |
| $R_n$ | — | — | $m_0 \div m_{17}$ | any | $m_0 \div m_{17}$ | any | — | — | flag at $n$ |
| $R_n$ | all | any | all | any | all | any | all | any | meta } optional |
| $R_n$ | all | any | all | any | all | any | all | any | meta |

TABLE XV
MULTISERVICE IMPLEMENTATION ALTERNATIVES
ADDRESSING PRECISE TIME SYNCHRONIZATION CHALLENGES

| Data Framing | Approach | $N_R$ @ flag | $E$ | Resolution |
|---|---|---|---|---|
| Cl. 146 delim. | fade-in at $n-1$ + flag at $n$ | Ev. Disp. Bit | $2^8$ | single/BT (exact) |
| Cl. 146 delim. | fade-in at $n-1$ + flag at $n$ | Ev. Disp. Bit | $2^8$ | single/(.5/$f_0$) (exact) |
| Cl. 146 delim. | fade-in + flag at $n$ + meta | Ev. Disp. Bit | $2^{>8}$ | ≥ single/BT |
| Cl. 146 delim. | fade-in + flag at $n$ + meta | Ev. INFO Bit | $2^{>8}$ | ≥ various/BT |

TABLE XVI
MULTISERVICE IMPLEMENTATION ALTERNATIVES
ADDRESSING IMMEDIATE PAYLOAD PREEMPTION CHALLENGES

| Approach | $N_{CoS}$ | $N_R$ @ flag | $E$ | Comment |
|---|---|---|---|---|
| Cl. 146 delim. + flag at $n$ | ≤ 3 | CoS Index | $2^{20}$ | 4th period acts as fade-in |
| Cl. 146 d. + fade-in + flag | > 3 | CoS Index | $2^{24}$ | explicit fade-in assembly |
| fade-in at $n-1$ + flag at $n$ | ≤ 9 | SSD/ESD | $2^8$ | max $N_{CoS}$ = min{9, 12} |
| fade-in + flag at $n$ + meta | any | SSD/ESD | $2^{>8}$ | meta at $n+1, +2, +3$, etc. |

TABLE XVII
MULTISERVICE IMPLEMENTATION ALTERNATIVES
ADDRESSING SIMULTANEOUS HETEROGENEOUS CHALLENGES

| Data Framing | Sync Carrier | Sync Signal | $N_{CoS}$ | Resolution | $E$ |
|---|---|---|---|---|---|
| Cl. 146 d. +FF | is · { fade-in + flag | pulse per span | = 2 | ≈ half/BT | $2^{24}$ |
| Cl. 146 d. +FFM | is · { Fi + Fl + meta | short message | > 2 | ≥ half/BT | $2^{>24}$ |
| fade-in + flag | same · fade-in + flag | pulse per span | = 2 | = single/$T_w$ | $2^8$ |
| Fi + Fl + meta | same · Fi + Fl + meta | short message | > 2 | > single/$T_w$ | $2^{>8}$ |

Among other, we do not restrict on the multiservice means to use the original delimitation, see Table X, or not.

Among other, we, however, restrict on that means to use a common pattern for any event insertion, see Table XIV briefly in conjunction with Table VIII again.[5]

## THE EXOTIC WAY

Insertion of an event, see Table XIV briefly again, generates an echo, see Tables XV, XVI, and XVII briefly, every time it is invoked by the coding means operating on the transmission channel, and, for the return, causes on the means to somehow neutralize that echo, e.g., attempting to perform a line of echo multiplexing rounds really canceling the echo, that in our case may be endless in time, however, see Table XI.

[5] $R_{n-1}$ numbers are applicable only in the case of an event flag at $n$.

Due to the obvious reason, a hopeful echo will be very rare and supported by a mocking one, see Table XII, for the most of time, therefore all we can securely count on is the available echo neutralization budget (area), see Table XIII.

However, if the real echo cancellation is our goal, it causes the use of something else besides a mocking echo.

We consider such a design of a multiservice coding means as being implemented the exotic way, i.e., unrealistic.

## THE PROGRESSIVE WAY

A multiservice may include a time synchronization service, characterized by a basic resolution, see Table XV, and a seamy preemption one, characterized by a set of classes of service,[6] see Table XVI, both the same time, see Table XVII.

[6] A modern IEEE 802.1D/Q bridge supports for up to 8 traffic classes.

TABLE XVIII
EXAMPLE MULTISERVICE OPERATION FRAGMENT

| Word Time Period, $n \rightarrow$ (shown event-arranged) | ⋯ | ⋯ | $n-5$ | $n-4$ | $n-3$ | $n-2$ | $n-1$ | $n$ | ⋯ | ⋯ |
|---|---|---|---|---|---|---|---|---|---|---|
| Channel State (Phase) actual in regular transmission modes | ⋯ | Data Streaming | Clause 146 Delimiting | Clause 146 Delimiting | Clause 146 Delimiting | Clause 146 Delimiting | Event Insertion | Event Insertion | Data Streaming | ⋯ |
| Coding Space Structure | ⋯ | $R_n$ | n/a | n/a | n/a | n/a | $R_{n-1}$ | $R_n$ | $R_n$ | ⋯ |
| Input Data Stream one to be multiplexed with other | ⋯ | Idle | SILENCE | SILENCE | DSPRST | SSD | Payload$_4$ | Payload$_5$ | Payload$_6$ | ⋯ |
| Input Sync Pulse Detection Scale (alternative A) when related with MII bit times | | | #1 #2 #3 #4 | #5 #6 #7 #8 | #9 #10 #11 #12 | #13 #14 #15 #16 | #17 #18 #19 #20 | #21 #22 #23 #24 | | |
| Selectable Pages from the respective dictionary | ⋯ | [1 2 3 4] | [— — — —] | [— — — —] | [1 2 3 4] | [1 — — 4] | [— 2 3 —] | [— 2 3 —] | [1 2 3 4] | ⋯ |
| Multiplexed Stream one to be transmitted on the line | ⋯ | Idle | SILENCE | SILENCE | DSPRST | SSD | **Ev. Fade-in** | **Event Flag** | Payload$_4$ (delayed) | ⋯ |
| Input Sync Pulse Detection Scale (alternative B) when related with transport letter half times | | | $1 ⋯ $6 | $7 ⋯ $12 | $13 ⋯ $18 | $19 ⋯ $24 | together provide a choice of 1 among 12×2 = 24 | | | |
| Echo Afore / Action Taken incl. echo-generating / Echo After | ⋯ | $\frac{1}{1}$ pass $\frac{1}{1}$ | $\frac{1}{1}$ pass $\frac{1}{1}$ | $\frac{1}{1}$ pass $\frac{1}{1}$ | $\frac{1}{1}$ pass $\frac{1}{1}$ | $\frac{1}{1}$ pass $\frac{1}{1}$ | $\frac{1}{1}$ mux $\frac{1}{2^4}$ | mux $\frac{1}{2^8}$ | mux $\frac{1}{2^8}$ | ⋯ |
| Letter Time Period, $t = 3n + \ldots \rightarrow$ | +0 +1 +2 | +0 +1 +2 | +0 +1 +2 | +0 +1 +2 | +0 +1 +2 | +0 +1 +2 | +0 +1 +2 | +0 +1 +2 | +0 +1 +2 | ⋯ |

TABLE XIX
CODING DELAYS

| Mixed Alt. | Syncronization Only | Preemption Only | TX/RX Delay | Unit |
|---|---|---|---|---|
| $E_\Sigma = V \times E$ | $E_\Sigma = V \times E + 16$ | $E_\Sigma = V \times E - 16$ | ceil ($E_\Sigma$ div 4) | $T_w$ |
| $V$ = max events per frame, $E$ = echo bits per event, see Tables XVII, XVI, XV, resp. | | | symmetric | 4·BT |

TABLE XX
LINGUISTIC MULTIPLEXING CHARACTER

| Technique $\rightarrow$ | Substitution | Representation |
|---|---|---|
| Application Subjects | any extra delimiters, event assemblies, mocking echo | bit shift, when $E \bmod 4 \neq 0$ nothing, when $E \bmod 4 = 0$ |

We consider such a design of a multiservice coding means as being implemented the progressive way.

## THE CONSERVATIVE WAY

A multiservice may include just one extra service over the original 10BASE-T1L duty, i.e., either of time, see Table XV again, or of data, see Table XVI again, but not both the same time, that simplifies the underlying coding means.

Anyway, at the heart of that service lies an event insertion pattern, see Table XIV, performing solo, or accompanying to the original delimitation, see Table X again, the proper choice of which is made in each particular case, carefully.

Anyway, the side effect of an event insertion is the echo it generates—accumulated with every event—we deal every next period with by a mocking round having the natural limits on its use within the frame, see Tables XII and XIII again.

With or without the delimitation, any event pattern consists at least of two assemblies, see Table XIV again, inserted into the multiplexed stream during the corresponding multiservice operation, see Table XVIII (example with delimitation).[7]

With or without the delimitation, any coding scheme results in reasonable coding delays, see Table XIX, whose variations, however, show as a little as no effect on the complexity of the underlying multiplexing process, see Table XX.

We consider such a design of a multiservice coding means as being implemented the conservative way.

## CONCLUSION

In this paper, we discussed a bunch of ways enriching the functionality of the 10BASE-T1L PCS,[8] all having the same ground, i.e., linguistic multiplexing applied over a dictionary of sufficient and redundant pages, behind them.

Among the ways, only the exotic one seems pure scientific, but interesting and important in theory, while the rest two are realistic enough to be implemented in practice.

Among the two, the progressive way introduces a synchronization service and a preemption service, both into and over the original duty of the considered Ethernet protocol.

Among the two, too, the conservative way limits the design with a single extra service enabled over the original duty.

---

[7]Corresponds to the options listed in the two upper rows of Table XV.

[8]This is the only entity we retrofit on, within the whole Physical Layer.